\documentclass[aps,prl,twocolumn,showpacs,10pt]{revtex4-1}

\usepackage{graphicx}
\usepackage[usenames,dvipsnames]{xcolor}
\usepackage{amsmath,amsfonts}
\usepackage{pdfpages}
\usepackage{hyperref}

\usepackage[normalem]{ulem}

\begin{document}
\title{Specialization and Bet-Hedging in Heterogeneous Populations}
\author{Steffen Rulands}
\author{David Jahn}
\author{Erwin Frey}

\affiliation{Department of Physics, Arnold-Sommerfeld-Center for Theoretical Physics and Center for NanoScience, Ludwig-Maximilians-Universit\"at M\"unchen, Theresienstrasse 37, D-80333 Munich, Germany}

\begin{abstract}
Phenotypic heterogeneity is a strategy commonly used by bacteria to rapidly adapt to changing environmental conditions. Here, we study the interplay between phenotypic heterogeneity and genetic diversity in spatially extended populations. By analyzing the spatio-temporal dynamics, we show that the level of mobility and the type of competition qualitatively influence the persistence of phenotypic heterogeneity. While direct competition generally promotes persistence of phenotypic heterogeneity, specialization dominates in models with indirect competition irrespective of the degree of mobility.
\end{abstract}

\maketitle

Genetic diversity and phenotypic heterogeneity are both commonly found in microbial and viral populations~\cite{Smits2006, Gefen2009, Rainey2011, Brockert2003, Porman2011, Rainey2011, Ramirez-Zavala2008, Soll1992}. However, in a homogeneous environment without differentiated niches, genetic diversity is difficult to maintain~\cite{Rainey1998}. Cyclic dominance has been identified as a factor promoting biodiversity in spatially extended systems~\cite{sinervo-1996-340, Durrett1997, Durrett:1998p203, Kerr2002, Kerr2007, Weber2014, Reichenbach2007, Reichenbach2007a}. For example,  bacterial model systems comprised of three genetically distinct strains of \emph{E. coli} exhibit three-strain coexistence in spatially extended homogeneous environments~\cite{Kerr2002, Weber2014}. 
In this system, a toxin-releasing strain kills a sensitive but not a resistant strain. The sensitive strain grows faster than the resistant strain which in turn grows faster than the toxin-producing strain. Recent theoretical studies have explored how demographic noise~\cite{traulsen-2005-95, Reichenbach2006, Reichenbach2007, Reichenbach2007a,  Claussen2008, Rulands2011,  Traulsen2012, Rulands2013a} or variability~\cite{Dobramysl2013, Dobramysl2013a}, mobility of individuals \cite{Reichenbach2007, Reichenbach2007a, Reichenbach2008}, as well as the topology of the food web~\cite{Knebel2013} and the interaction network~\cite{Szabo2004} affect maintenance of genotypic diversity. All of these studies assume that genotypes are linked to a single phenotype. However, some bacteria use a bet-hedging strategy, stochastically switching between different phenotypic states to minimize the risk of population extinction, e.g. during exposure to antibiotics~\cite{Balaban2004, Gefen2009}. Switching between cyclically dominating phenotypes in \emph{E.coli} can be experimentally realized using synthetic genetic switches, which lead to stochastic switching between toxin production, immunity and sensitivity~\cite{Gardner2000}. Is phenotypic heterogeneity maintained under these conditions or does specialization pervail, and what is the role of mobility and the interaction between individuals?

\begin{figure}[htb]
\includegraphics[width=0.9\columnwidth]{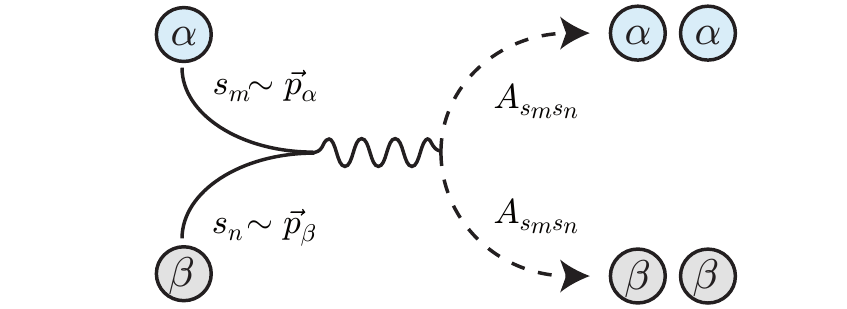}
\caption{
\emph{Illustration of competition in the heterogeneous ecological model.} When individuals with genotypes $\alpha$ and $\beta$ engage, their phenotypic states $s_m$ and $s_n$ are chosen randomly according to a probability distribution. The outcome of the reaction is specified by the  interaction matrix ${A}_{s_m  s_n}$.
}
\label{fig:model}
\end{figure}

We address these questions by studying the dynamics of spatially extended populations which initially contain $N$ individuals of $G$ different genotypes. Each of these genotypes $\alpha \in \{ 1, \ldots, G \} $ is defined by its degree of phenotypic heterogeneity, \emph{i.e.}\/ a set of probabilities  $\vec{p}_\alpha=\left(p_\alpha^1,\dots,p_\alpha^M\right)$ with $p_\alpha^m$ signifying the probability that a genotype $\alpha$ is in a particular phenotypic state $s_m \in \{ s_1,\ldots, s_M \}$ (e.g. capable of producing immunity proteins) at the moment of interaction with another genotype $\beta$, cf. Fig.~\ref{fig:model}.  For specificity, we will focus on systems with $M=3$ phenotypic states and defer a discussion of a larger number of states to the Supplementary Material (SM)~\cite{SM}. Then, the phenotypes $s_m$ may, for example, refer to one of the three traits of \emph{E. coli} discussed above. We consider two distinct ecological scenarios, where, as in the \emph{E.coli} model system, phenotype $s_m$ outcompetes phenotype $s_{m+1}$ cyclically. In the first class of models, termed Lotka-Volterra (LV) models~\cite{lotka1920,volterra-1926-31}, selection and reproduction occur simultaneously, in that competition is combined into a single event where the competition between two individuals leads to the immediate replacement of the weaker by the stronger individual:  $I + J \to I + I$. LV models mimic predator-prey interactions and they are applicable to situations in which competition is not limited by the availability of resources, such as nutrients on an agar plate. They have, for example, been used to study beneficial mutations in growing bacterial colonies~\cite{Lehe2012} or spatial competition in strains of budding yeast~\cite{Korolev2012}. In the second class of models, originally proposed by May and Leonard (ML)~\cite{May1975}, selection and reproduction are two separate processes. An interaction between two individuals with different phenotypes leads to the death of the weaker phenotype and makes resources available: $I + J \to I + \emptyset$.  Reproduction then follows as a second process which recolonizes this empty space: $ I + \emptyset \to I + I$. In an ecological context, these empty sites effectively introduce the factor `carrying capacity' and thus mimic the effects of resource limitation. ML models have been employed to model synthetic \emph{E. coli} systems~\cite{Kerr2002,Weber2014}. In both models, the genotype $\alpha_i$ of individual $I$ is transmitted to its offspring. 

In the well-mixed case both models possess a fixed point given by an equal abundance of each genotype, as well as $N$ absorbing states, corresponding to the extinction of all but one genotype. However, the nonlinear dynamics in both models is vastly different: The LV model shows a maximum number of conserved quantities corresponding to neutrally stable, closed orbits in the space of genotype abundancies~\cite{Knebel2013}. By contrast, the ML model shows heteroclinic orbits emerging from trajectories spiraling out from an unstable reactive fixed point~\cite{SM}.

In this Letter we show that the degree of mobility and the type of competition qualitatively influence the loss of genetic diversity, and that each of these factors has a major impact on the persistence of phenotypic heterogeneity. For direct competition, as in LV models, the evolutionary outcome strongly depends on the mobility. We find that in well-mixed populations phenotypic heterogeneity is favored, whereas spatial correlations promote unique phenotypes at low mobility levels. By contrast, if competition is mediated by the limited availability of resources as in the ML model, phenotypic heterogeneity is lost irrespective of the degree of mobility.

Specifically, we study a lattice gas model where at a given time $t$ the state $\mathcal{C}$ of the population is characterized by a set of genotypes $\vec{p}_{\alpha_i}$ and lattice positions ${\bf{r}}_i(t)$ for each individual $i \in \{ 1, \ldots, N\} $: $\mathcal{C}(t)= \{ \alpha_i, {\bf r}_i (t) \}_{i=1,\dots,N}$.  Each lattice site on a two-dimensional square lattice with $L^2$ sites is occupied by at most one individual. The linear dimension of the lattice is taken as the basic length unit. When two neighboring individuals interact each randomly chooses a phenotype according to its respective probability vector. The outcome of these pairwise competitions is described in terms of an interaction matrix ${\bf A}$, whose entries $A_{s s'}$ denote the rate at which phenotype $s$ outcompetes phenotype $s'$. For simplicity, we choose a symmetric model, where all finite rates are the same, and equal to $1$ to fix the time scale~\footnote{This can easily be generalized to asymmetric competition between phenotypes by considering phenotype dependent competition rates.}. Mobility of individuals is implemented as a nearest-neighbor exchange process at a rate $\epsilon$, $I  + J \to J  + I$, where $I$ and $J$ denote individuals or empty spaces $\emptyset$. Macroscopically this exchange process leads to diffusion with an effective diffusion constant $D \! = \! \epsilon / (2L^2)$~\cite{Reichenbach2007}. In dimensionless units $D$ gives the mean-square displacement of a particle between two reactions.
  
\begin{figure}[t]
\includegraphics[width=\columnwidth]{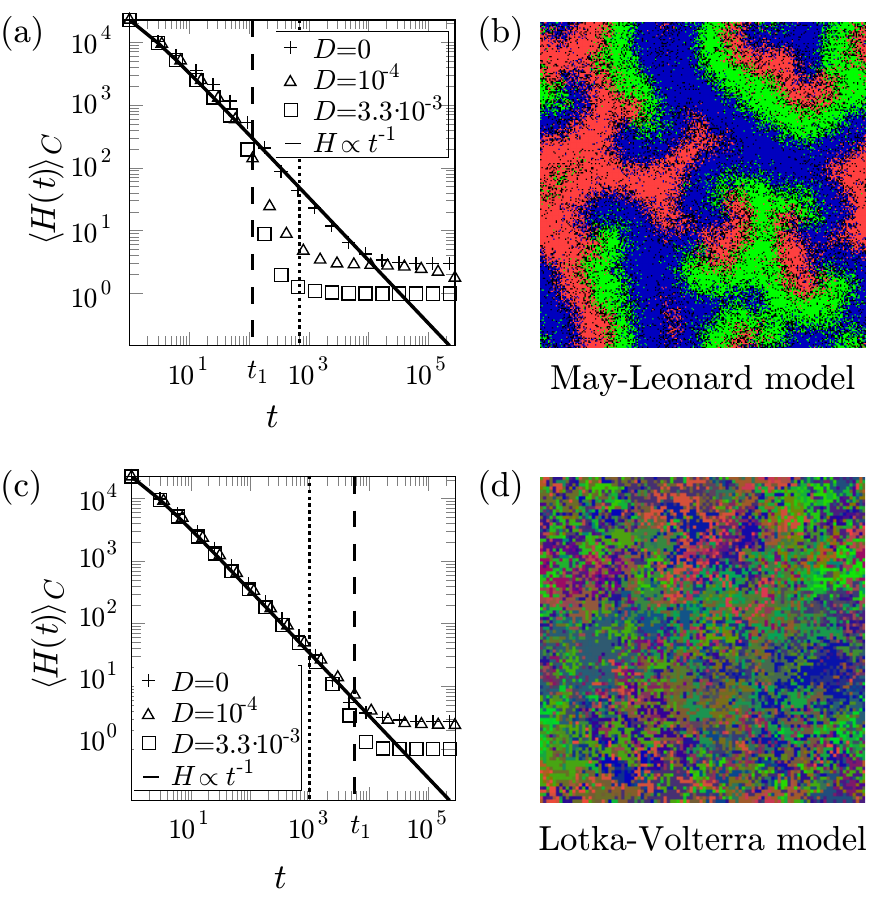}
\caption{
(a) Time evolution of genetic diversity for the ML model. 
(b) A typical configuration of the ML model for large times [$t=660$, dotted line in (a)] and a low value of the diffusion constant ($D=10^{-4}$). Different colors (gray scales) signify the probability to be in any of three phenotypic states: red (light gray), green (medium gray) or blue (dark gray) denotes a high probability to be in the phenotypic states $s_1$, $s_2$ or $s_3$, respectively. 
(c) Time evolution of genetic diversity for the LV model.
(d) A typical configuration of the LV model for large times [$t=10^3$, dotted line in (c), $L=100$].
}
\label{fig:hets}
\end{figure}

We performed stochastic simulations of both classes of ecological models employing periodic boundary conditions and a sequential updating algorithm. All simulations were started from an initial state comprising $G$ genotypes chosen randomly according to a uniform distribution on the unit simplex $\Delta^2$, and then distributed randomly over the lattice. As time progresses, competition between these genotypes reduces genetic diversity in the population: Figures~\ref{fig:hets} (a,c) show the number of different genotypes, $H(t)$, averaged over $10^5$ (a) and $5\cdot 10^4$ (c) realizations $\mathcal{C}$. 
Concomitant with the loss of genetic diversity, spatio-temporal patterns and correlations emerge. While for large $D$ both models quickly reach a state where only one genotype is left in the population, for small $D$ there are long-lived metastable states containing three distinct genotypes [Figs.~\ref{fig:hets} (a, c)]. 

\begin{figure*}[t!]
\includegraphics[width=\textwidth]{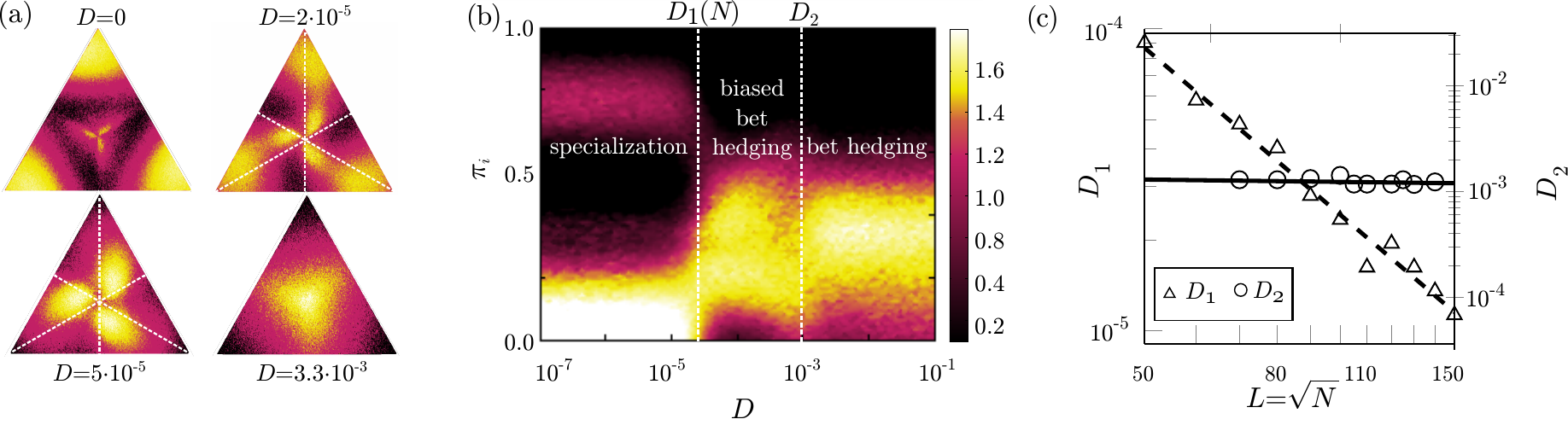}
\caption{
\emph{Asymptotic genotypes for the heterogeneous LV model.} 
(a) Probability density $\mathcal{P}_\infty (\vec{\pi})$ of the asymptotic genotypes $\vec{\pi}$ for different values of the diffusion coefficient $D$. Color (gray scale) denotes the value of $\mathcal{P}_\infty$, such that white signifies a high value and black a low value. The maxima of $\mathcal{P}_\infty$ identify successful genotypes. 
(b) Marginal probability distribution to be in any of the components of $\vec{\pi}$. We identify two threshold values, $D_1(N)$ and $D_2$, separating distinct outcomes of the evolutionary dynamics as indicated in the graph ($L=80$). 
(c) Scaling of the threshold values with system size $L = \sqrt{N}$: $D_1\propto 1/N$ and $D_2=\text{const.}$. 
}
\label{fig:asystrats}
\end{figure*}

Initially, quite independent of the value for $D$ and the class of ecological model, we observe $\langle H(t) \rangle_{\mathcal{C}}\propto t^{-1}$. This is because genetic diversity is high and, therefore, selection occurs irrespective of the genotype: Loss of genetic diversity is then described by a neutral coalescence process; the rate  is given by the probability that the two competing individuals are in distinct phenotypic states, $k=2/3$.  Fluctuations can be neglected and the dynamics of this process can be described in terms of mean-field kinetics, with $\partial_t H = -k H^2$, and integration yields $H(t) = N/(1 + k t)$ in good agreement with our numerical results [Figs.~\ref{fig:hets} (a, c)]. As time proceeds and genetic diversity decreases, spatio-temporal patterns form and correlations emerge [Figs.~\ref{fig:hets}(b,d)]. As a consequence, the neutral regime ends at some characteristic time $t_1$, and thereafter the genealogical dynamics is driven by evolutionary forces, \emph{i.e.} success in reproduction depends on how each genotype interacts with its neighbors. We observe that while for the ML model $t_1$ scales logarithmically with the population size, $t_1\propto \ln N$, it scales linearly for the LV model, $t_1\propto N$; see Supplemental Material~\cite{SM}. This is due to the nature of the respective orbits in phase space~\cite{frey2010}: In the ML model heteroclinic orbits generate a drift towards the phase space boundary, such that the ensuing extinction process is exponentially accelerated, which results in logarithmic scaling. In contrast, the phase portrait of the LV model exhibits neutrally stable orbits, and the stochastic dynamics performs an unbiased random walk~\cite{Knebel2013}.
This implies $t_1\propto N$, and thereby fixation occurs on a larger time scale. We also find that the rate of decrease of genetic diversity changes with the diffusion constant, most prominently in the ML model: The smaller $D$ the slower the extinction of genetic diversity. Hence, spatial structures not only stabilize systems of cyclically interacting species~\cite{Reichenbach2007, Reichenbach2008, Reichenbach2008b, Rulands2011, Rulands2013a}, but also promote genetic diversity therein. The reason for this remarkable behavior is that spatial structures consist of genetically identical individuals. Reactions between different genotypes, therefore, only occur at domain boundaries and thereby globally at a lower rate.

As time progresses spatial structures become more pronounced and genetic heterogeneity reaches a stationary level [Fig.~\ref{fig:hets}]. We find two qualitatively different regimes:  For low $D$, we observe a metastable state comprised of three distinct genotypes. This transient biodiversity is maintained by spatial alliances of individuals with identical genotypes, resulting in spiral waves (ML) or strong spatial correlations (LV), as previously studied for competing species with pure strategies~\cite{Reichenbach2006, Reichenbach2007, Reichenbach2007a, Reichenbach2008, Reichenbach2008b, Claussen2008,  Rulands2011, Rulands2013a,Szczesny2013}. By contrast, for large $D$, the population ends up in one of the absorbing states corresponding to the extinction of all but one genotype. We refer to those states as \emph{asymptotic genotype} $\vec{\pi}$. Which genotype becomes dominant under what conditions, and how is this affected by the kind of competition between individuals? To answer these questions we consider many realizations $\mathcal{C}$ of the population dynamics and determine the probability density $\mathcal{P}_\infty(\vec{\pi})$ of asymptotic genotypes on the simplex $\vec{\pi}\in\Delta^2$ [Figs.~\ref{fig:asystrats}(a,b)]. $\Delta^2$ is also called a \emph{Pareto front}, i.e.\ the set of all Pareto-optimal strategies in response to three conflicting objectives given by the environment. While previous work mainly concerned with the distribution of strategies in stationary environments~\cite{Sheftel2013}, we here study how these strategies dynamically distribute in response to objectives given by the local composition of the population.
Maxima of $\mathcal{P}_\infty$ identify the evolutionarily most successful genotypes~\footnote{Our simulations show that the three surviving genotypes in the metastable regime contribute with equal probability to the asymptotic states. For the reason of numerical efficiency we therefore computed $P_\infty(\vec{\pi})$ at times corresponding to the metastable regime.}. 

We start the discussion with the LV model, cf. Fig.~\ref{fig:asystrats}. Our simulations show that, which genotype is evolutionarily most successful strongly depends on the mobility, and one can identify three distinct regimes: If diffusion is slow, it is evolutionarily most advantageous to \emph{specialize}, \emph{i.e.}\/ to adopt and retain any one of the three phenotypes; $P_\infty$ is largest in the corners of the simplex. In contrast, for large $D$, the most successful individuals are \emph{bet-hedgers}, \emph{i.e.}\/  genotypes with nearly equal probabilities for each of the three phenotypes. For intermediate values of $D$, the most successful individuals adopt a bet-hedging strategy that is \emph{biased} towards one of the three phenotypes. The boundaries between these three qualitatively different regimes,  $D_1$ and $D_2$, are clearly visible in Fig.~\ref{fig:asystrats}(b), which shows the marginal probability distribution for each of the three components of $\vec{\pi}$~\footnote{For symmetry reasons all three of these marginal distributions are identical.}. Beyond that, the threshold $D_2$ also separates neutrally stable from metastable dynamics and therefore marks a sharp transition in the first passage times to any of the absorbing states.

For fast diffusion, $D>D_2$, the characteristic length scale of spatial patterns is larger than the system size, and, therefore, the dynamics is effectively that of a well-mixed system~\cite{Reichenbach2007,Reichenbach2008b}. Then the interaction between individuals with two different genotypes is well described by a mean-field approximation: the probability that an individual of genotype $\alpha$ outcompetes one of genotype $\beta$  is given by $w_{\alpha \beta} = {\vec p}_\alpha^\mathrm{\, T} \text{\bf{A}} {\vec p}_\beta^\mathrm{} = p^1_\alpha p^2_\beta + p^2_\alpha p^3_\beta + p^3_\alpha p^1_\beta$. This implies a net transition rate between genotypes, $W_{\alpha \beta} = w_{\alpha \beta}-w_{\beta\alpha}$, such that the fraction $x_\alpha$ of individuals with genotype $\alpha$ obeys the rate equation:
\begin{equation}
  \partial_t x_\alpha (t) = x_\alpha (t) \sum^{3}_{\beta = 1} W_{\alpha \beta} \, x_\beta (t) \, .
  \label{eq:replicator}
\end{equation}
Since $W_{\alpha \beta} $ is a skew-symmetric matrix, this corresponds to the replicator equation of a $G$-species conservative LV model, whose dynamics has recently been classified~\cite{Knebel2013}. Obviously, a strictly bet-hedging strategy with $\vec p_B = (\frac13, \frac13, \frac13)$ is a fixed-point of Eq.~(\ref{eq:replicator}); since $W_{B \beta} = 0$ it can not be outcompeted by any other genotype $\beta$. Moreover, the particular form of $W_{\alpha \beta} $ implies that all orbits are neutrally stable and periodic~\cite{Hofbauer1998}. Since the bet-hedging genotype, $\vec p_B$, is furthest away from the boundaries of the simplex, the corresponding mean first passage time into the absorbing states is the longest~\cite{Reichenbach2006, Parker2009, Dobrinevski2012}. Hence, for large times,  bet-hedging genotypes are the most abundant  [Fig.~\ref{fig:asystrats}(a), bottom right].

With decreasing diffusion constant $D$ the hopping rate $\epsilon$ between neighboring lattice sites eventually becomes much smaller than the reactions rates, $\epsilon \ll 1$. This defines a threshold for $D$ which should scale as $D_1 \sim 1/N$, as confirmed by our simulations [Fig.~\ref{fig:asystrats} (c)]. For $D<D_1$, the dynamics is reaction-dominated and, therefore, a domain boundary between two different genotypes advances mainly due to competitive takeover and not due to hopping between neighboring lattice sites. This leads to rather smooth domain boundaries, which move at a speed proportional to the net transition rate $W_{\alpha \beta}$.  This invasion speed is highest, if either genotype is a specialist. Hence, while specialists invade other genotypes fastest, they also are also most susceptible to displacement by other genotypes. This makes it difficult to see who will eventually win the race. The decisive factor is that the initial coarsening process leads to spatial domains consisting of selectively neutral genotypes which, in addition, are spatially organized such that fast advancing specialists form a strategic alliance with generalists, who are able to defend the territory, because they are intrinsically more resistant to invasion~\cite{SM}. Those profiting the most from this alliance are the specialists since it enables them to invade new territory fast. Hence by a `first come first served' principle, specialized genotypes outcompete their bet-hedging counterparts, and, for large times, the dynamics shows (transient) cyclic competition between three specialized genotypes [Fig.~\ref{fig:asystrats}(b)]. 

Interestingly, we also find an intermediate parameter regime, $D_1 < D < D_2$, where the dynamics shows prolonged metastable states. Unlike the specialists observed for $D<D_1$, the surviving genotypes now partly favor one particular phenotype, but retain a non-negligible propensity to adopt the other phenotypic states [propeller-like structure in Fig.~\ref{fig:asystrats}(a), bottom left]. Since now nearest neighbour exchange processes occur at the same time scale as competitive interactions the domain boundaries are fuzzy. Moreover, due to an increasing mean path length associated with $D$, domains are frequently intruded by particles with a distinct genotype. As a result, the surviving genotypes are characterized by a trade-off between invasion speed, given by $W_{\alpha \beta}$ and robustness against hostile invasion, given by a broad distribution of phenotypic states. A more detailed discussion is given in the SM~\cite{SM}.

For the ML model, we find a remarkably different behavior. There, independent of the value of the diffusion constant $D$, the population is asymptotically dominated by specialists~\cite{SM}. Phenotypic heterogeneity does not provide an evolutionary advantage in a setting, where limited resources lead to indirect competition. The dynamics asymptotically approaches  the classical ML model~\cite{Reichenbach2007, Reichenbach2008, Rulands2011, Rulands2013a}, as is demonstrated in the SM~\cite{SM}.

In conclusion, we have investigated the spatio-temporal dynamics of heterogeneous populations with an initially high degree of genetic diversity where individuals show a varying degree of phenotypic heterogeneity. We have found that
 the degree of mobility, as well as the type of competition, qualitatively affect both the loss of genetic diversity and the maintenance of phenotypic heterogeneity. In the LV model, the degree of phenotypic heterogeneity changes qualitatively at certain threshold values of the diffusion constant. In contrast to this behavior, in the ML model specialists always dominate the population in the long run. For heterogeneous bacterial populations this means that the survival of phenotypic heterogeneity depends both on the degree of mixing and the relative availability of nutrients. The impact of mobility and the type of competition on the survival of phenotypic heterogeneity is not restricted to these models. In fact, we think that the mechanisms behind these phenomena are generic, in the sense that they only rely on basic properties of the underlying nonlinear dynamics, namely neutrally stable orbits as in LV models or heteroclinic cycles as in ML models. This view is supported by the fact that we observed the same behavior in a more complex model with four species~\cite{SM, Drobrinevski2014}. We therefore believe that our findings apply to a broad class of ecological contexts. While we have reported results for one~\cite{SM}, two and infinite spatial dimensions the dynamics in three dimensions remains an open question for future research.

\begin{acknowledgments}
This research was supported by the German Excellence Initiative via the program  `NanoSystems Initiative Munich' and the Deutsche Forschungsgemeinschaft via 
the Priority Programme "Phenotypic heterogeneity and sociobiology of bacterial populations" (SPP 1617). S.R. gratefully acknowledges support of the Wellcome Trust (grant number 098357/Z/12/Z). We thank Alejandro Zielinski, Johannes Knebel and Markus Weber for fruitful and stimulating discussions.
\end{acknowledgments}


\begin{thebibliography}{46}%
\makeatletter
\providecommand \@ifxundefined [1]{%
 \@ifx{#1\undefined}
}%
\providecommand \@ifnum [1]{%
 \ifnum #1\expandafter \@firstoftwo
 \else \expandafter \@secondoftwo
 \fi
}%
\providecommand \@ifx [1]{%
 \ifx #1\expandafter \@firstoftwo
 \else \expandafter \@secondoftwo
 \fi
}%
\providecommand \natexlab [1]{#1}%
\providecommand \enquote  [1]{``#1''}%
\providecommand \bibnamefont  [1]{#1}%
\providecommand \bibfnamefont [1]{#1}%
\providecommand \citenamefont [1]{#1}%
\providecommand \href@noop [0]{\@secondoftwo}%
\providecommand \href [0]{\begingroup \@sanitize@url \@href}%
\providecommand \@href[1]{\@@startlink{#1}\@@href}%
\providecommand \@@href[1]{\endgroup#1\@@endlink}%
\providecommand \@sanitize@url [0]{\catcode `\\12\catcode `\$12\catcode
  `\&12\catcode `\#12\catcode `\^12\catcode `\_12\catcode `\%12\relax}%
\providecommand \@@startlink[1]{}%
\providecommand \@@endlink[0]{}%
\providecommand \url  [0]{\begingroup\@sanitize@url \@url }%
\providecommand \@url [1]{\endgroup\@href {#1}{\urlprefix }}%
\providecommand \urlprefix  [0]{URL }%
\providecommand \Eprint [0]{\href }%
\providecommand \doibase [0]{http://dx.doi.org/}%
\providecommand \selectlanguage [0]{\@gobble}%
\providecommand \bibinfo  [0]{\@secondoftwo}%
\providecommand \bibfield  [0]{\@secondoftwo}%
\providecommand \translation [1]{[#1]}%
\providecommand \BibitemOpen [0]{}%
\providecommand \bibitemStop [0]{}%
\providecommand \bibitemNoStop [0]{.\EOS\space}%
\providecommand \EOS [0]{\spacefactor3000\relax}%
\providecommand \BibitemShut  [1]{\csname bibitem#1\endcsname}%
\let\auto@bib@innerbib\@empty
\bibitem [{\citenamefont {Smits}\ \emph {et~al.}(2006)\citenamefont {Smits},
  \citenamefont {Kuipers},\ and\ \citenamefont {Veening}}]{Smits2006}%
  \BibitemOpen
  \bibfield  {author} {\bibinfo {author} {\bibfnamefont {W.~K.}\ \bibnamefont
  {Smits}}, \bibinfo {author} {\bibfnamefont {O.~P.}\ \bibnamefont {Kuipers}},
  \ and\ \bibinfo {author} {\bibfnamefont {J.-W.}\ \bibnamefont {Veening}},\
  }\href {\doibase 10.1038/nrmicro1381} {\bibfield  {journal} {\bibinfo
  {journal} {Nat. Rev. Microbiol.}\ }\textbf {\bibinfo {volume} {4}},\ \bibinfo
  {pages} {259} (\bibinfo {year} {2006})}\BibitemShut {NoStop}%
\bibitem [{\citenamefont {Gefen}\ and\ \citenamefont
  {Balaban}(2009)}]{Gefen2009}%
  \BibitemOpen
  \bibfield  {author} {\bibinfo {author} {\bibfnamefont {O.}~\bibnamefont
  {Gefen}}\ and\ \bibinfo {author} {\bibfnamefont {N.~Q.}\ \bibnamefont
  {Balaban}},\ }\href {\doibase 10.1111/j.1574-6976.2008.00156.x} {\bibfield
  {journal} {\bibinfo  {journal} {FEMS Microbiol. Rev.}\ }\textbf {\bibinfo
  {volume} {33}},\ \bibinfo {pages} {704} (\bibinfo {year} {2009})}\BibitemShut
  {NoStop}%
\bibitem [{\citenamefont {Rainey}\ \emph {et~al.}(2011)\citenamefont {Rainey},
  \citenamefont {Beaumont}, \citenamefont {Ferguson}, \citenamefont {Gallie},
  \citenamefont {Kost}, \citenamefont {Libby},\ and\ \citenamefont
  {Zhang}}]{Rainey2011}%
  \BibitemOpen
  \bibfield  {author} {\bibinfo {author} {\bibfnamefont {P.~B.}\ \bibnamefont
  {Rainey}}, \bibinfo {author} {\bibfnamefont {H.~J.~E.}\ \bibnamefont
  {Beaumont}}, \bibinfo {author} {\bibfnamefont {G.~C.}\ \bibnamefont
  {Ferguson}}, \bibinfo {author} {\bibfnamefont {J.}~\bibnamefont {Gallie}},
  \bibinfo {author} {\bibfnamefont {C.}~\bibnamefont {Kost}}, \bibinfo {author}
  {\bibfnamefont {E.}~\bibnamefont {Libby}}, \ and\ \bibinfo {author}
  {\bibfnamefont {X.-X.}\ \bibnamefont {Zhang}},\ }\href@noop {} {\bibfield
  {journal} {\bibinfo  {journal} {Microbial Cell Factories}\ }\textbf {\bibinfo
  {volume} {10}},\ \bibinfo {pages} {S14} (\bibinfo {year} {2011})}\BibitemShut
  {NoStop}%
\bibitem [{\citenamefont {Brockert}\ \emph {et~al.}(2003)\citenamefont
  {Brockert}, \citenamefont {Lachke}, \citenamefont {Srikantha}, \citenamefont
  {Pujol}, \citenamefont {Galask},\ and\ \citenamefont {Soll}}]{Brockert2003}%
  \BibitemOpen
  \bibfield  {author} {\bibinfo {author} {\bibfnamefont {P.~J.}\ \bibnamefont
  {Brockert}}, \bibinfo {author} {\bibfnamefont {S.~A.}\ \bibnamefont
  {Lachke}}, \bibinfo {author} {\bibfnamefont {T.}~\bibnamefont {Srikantha}},
  \bibinfo {author} {\bibfnamefont {C.}~\bibnamefont {Pujol}}, \bibinfo
  {author} {\bibfnamefont {R.}~\bibnamefont {Galask}}, \ and\ \bibinfo {author}
  {\bibfnamefont {D.~R.}\ \bibnamefont {Soll}},\ }\href {\doibase
  10.1128/IAI.71.12.7109} {\bibfield  {journal} {\bibinfo  {journal} {Infect
  Immun.}\ }\textbf {\bibinfo {volume} {71}},\ \bibinfo {pages} {7109}
  (\bibinfo {year} {2003})}\BibitemShut {NoStop}%
\bibitem [{\citenamefont {Porman}\ \emph {et~al.}(2011)\citenamefont {Porman},
  \citenamefont {Alby}, \citenamefont {Hirakawa},\ and\ \citenamefont
  {Bennett}}]{Porman2011}%
  \BibitemOpen
  \bibfield  {author} {\bibinfo {author} {\bibfnamefont {A.~M.}\ \bibnamefont
  {Porman}}, \bibinfo {author} {\bibfnamefont {K.}~\bibnamefont {Alby}},
  \bibinfo {author} {\bibfnamefont {M.~P.}\ \bibnamefont {Hirakawa}}, \ and\
  \bibinfo {author} {\bibfnamefont {R.~J.}\ \bibnamefont {Bennett}},\ }\href
  {\doibase 10.1073/pnas.1112076109} {\bibfield  {journal} {\bibinfo  {journal}
  {Proc. Nat. Acad. Sci. USA}\ }\textbf {\bibinfo {volume} {108}},\ \bibinfo
  {pages} {21158} (\bibinfo {year} {2011})}\BibitemShut {NoStop}%
\bibitem [{\citenamefont {Ram\'{\i}rez-Zavala}\ \emph
  {et~al.}(2008)\citenamefont {Ram\'{\i}rez-Zavala}, \citenamefont {Reuss},
  \citenamefont {Park}, \citenamefont {Ohlsen},\ and\ \citenamefont
  {Morschh\"{a}user}}]{Ramirez-Zavala2008}%
  \BibitemOpen
  \bibfield  {author} {\bibinfo {author} {\bibfnamefont {B.}~\bibnamefont
  {Ram\'{\i}rez-Zavala}}, \bibinfo {author} {\bibfnamefont {O.}~\bibnamefont
  {Reuss}}, \bibinfo {author} {\bibfnamefont {Y.-N.}\ \bibnamefont {Park}},
  \bibinfo {author} {\bibfnamefont {K.}~\bibnamefont {Ohlsen}}, \ and\ \bibinfo
  {author} {\bibfnamefont {J.}~\bibnamefont {Morschh\"{a}user}},\ }\href
  {\doibase 10.1371/journal.ppat.1000089} {\bibfield  {journal} {\bibinfo
  {journal} {PLoS Pathog.}\ }\textbf {\bibinfo {volume} {4}},\ \bibinfo {pages}
  {e1000089} (\bibinfo {year} {2008})}\BibitemShut {NoStop}%
\bibitem [{\citenamefont {Soll}(1992)}]{Soll1992}%
  \BibitemOpen
  \bibfield  {author} {\bibinfo {author} {\bibfnamefont {D.~R.}\ \bibnamefont
  {Soll}},\ }\href
  {http://www.pubmedcentral.nih.gov/articlerender.fcgi?artid=358234\&tool=pmcentrez\&rendertype=abstract}
  {\bibfield  {journal} {\bibinfo  {journal} {Clin. Microbiol. Rev.}\ }\textbf
  {\bibinfo {volume} {5}},\ \bibinfo {pages} {183} (\bibinfo {year}
  {1992})}\BibitemShut {NoStop}%
\bibitem [{\citenamefont {Rainey}\ and\ \citenamefont
  {Travisano}(1998)}]{Rainey1998}%
  \BibitemOpen
  \bibfield  {author} {\bibinfo {author} {\bibfnamefont {P.~B.}\ \bibnamefont
  {Rainey}}\ and\ \bibinfo {author} {\bibfnamefont {M.}~\bibnamefont
  {Travisano}},\ }\href@noop {} {\bibfield  {journal} {\bibinfo  {journal}
  {Nature}\ }\textbf {\bibinfo {volume} {394}},\ \bibinfo {pages} {69}
  (\bibinfo {year} {1998})}\BibitemShut {NoStop}%
\bibitem [{\citenamefont {Sinervo}\ and\ \citenamefont
  {Lively}(1996)}]{sinervo-1996-340}%
  \BibitemOpen
  \bibfield  {author} {\bibinfo {author} {\bibfnamefont {B.}~\bibnamefont
  {Sinervo}}\ and\ \bibinfo {author} {\bibfnamefont {C.~M.}\ \bibnamefont
  {Lively}},\ }\href
  {http://www.nature.com/nature/journal/v380/n6571/abs/380240a0.html
  papers://6bd6bfca-595b-4d6b-94ac-3305a1243fe1/Paper/p815} {\bibfield
  {journal} {\bibinfo  {journal} {Nature}\ }\textbf {\bibinfo {volume} {380}},\
  \bibinfo {pages} {240} (\bibinfo {year} {1996})}\BibitemShut {NoStop}%
\bibitem [{\citenamefont {Durrett}\ and\ \citenamefont
  {Levin}(1997)}]{Durrett1997}%
  \BibitemOpen
  \bibfield  {author} {\bibinfo {author} {\bibfnamefont {R.}~\bibnamefont
  {Durrett}}\ and\ \bibinfo {author} {\bibfnamefont {S.}~\bibnamefont
  {Levin}},\ }\href {\doibase 10.1006/jtbi.1996.0292} {\bibfield  {journal}
  {\bibinfo  {journal} {J. Theor. Biol.}\ }\textbf {\bibinfo {volume} {185}},\
  \bibinfo {pages} {165} (\bibinfo {year} {1997})}\BibitemShut {NoStop}%
\bibitem [{\citenamefont {Durrett}\ and\ \citenamefont
  {Levin}(1998)}]{Durrett:1998p203}%
  \BibitemOpen
  \bibfield  {author} {\bibinfo {author} {\bibfnamefont {R.}~\bibnamefont
  {Durrett}}\ and\ \bibinfo {author} {\bibfnamefont {S.}~\bibnamefont
  {Levin}},\ }\href@noop {} {\bibfield  {journal} {\bibinfo  {journal} {Theor.
  Pop. Biol.}\ }\textbf {\bibinfo {volume} {53}},\ \bibinfo {pages} {30}
  (\bibinfo {year} {1998})}\BibitemShut {NoStop}%
\bibitem [{\citenamefont {Kerr}\ \emph {et~al.}(2002)\citenamefont {Kerr},
  \citenamefont {Riley}, \citenamefont {Feldman},\ and\ \citenamefont
  {Bohannan}}]{Kerr2002}%
  \BibitemOpen
  \bibfield  {author} {\bibinfo {author} {\bibfnamefont {B.}~\bibnamefont
  {Kerr}}, \bibinfo {author} {\bibfnamefont {M.~A.}\ \bibnamefont {Riley}},
  \bibinfo {author} {\bibfnamefont {M.~W.}\ \bibnamefont {Feldman}}, \ and\
  \bibinfo {author} {\bibfnamefont {B.~J.}\ \bibnamefont {Bohannan}},\ }\href
  {http://eutils.ncbi.nlm.nih.gov/entrez/eutils/elink.fcgi?cmd=prlinks\&dbfrom=pubmed\&retmode=ref\&id=12110887
  papers://6bd6bfca-595b-4d6b-94ac-3305a1243fe1/Paper/p185} {\bibfield
  {journal} {\bibinfo  {journal} {Nature}\ }\textbf {\bibinfo {volume} {418}},\
  \bibinfo {pages} {171} (\bibinfo {year} {2002})}\BibitemShut {NoStop}%
\bibitem [{\citenamefont {Kerr}(2007)}]{Kerr2007}%
  \BibitemOpen
  \bibfield  {author} {\bibinfo {author} {\bibfnamefont {B.}~\bibnamefont
  {Kerr}},\ }in\ \href@noop {} {\emph {\bibinfo {booktitle} {Bacteriocins}}},\
  \bibinfo {editor} {edited by\ \bibinfo {editor} {\bibfnamefont {M.~A.}\
  \bibnamefont {Riley}}\ and\ \bibinfo {editor} {\bibfnamefont {M.~A.}\
  \bibnamefont {Chavan}}}\ (\bibinfo  {publisher} {Springer},\ \bibinfo
  {address} {New York},\ \bibinfo {year} {2007})\ pp.\ \bibinfo {pages}
  {111--134}\BibitemShut {NoStop}%
\bibitem [{\citenamefont {Weber}\ \emph {et~al.}(2014)\citenamefont {Weber},
  \citenamefont {Poxleitner}, \citenamefont {Hebisch}, \citenamefont {Frey},\
  and\ \citenamefont {Opitz}}]{Weber2014}%
  \BibitemOpen
  \bibfield  {author} {\bibinfo {author} {\bibfnamefont {M.~F.}\ \bibnamefont
  {Weber}}, \bibinfo {author} {\bibfnamefont {G.}~\bibnamefont {Poxleitner}},
  \bibinfo {author} {\bibfnamefont {E.}~\bibnamefont {Hebisch}}, \bibinfo
  {author} {\bibfnamefont {E.}~\bibnamefont {Frey}}, \ and\ \bibinfo {author}
  {\bibfnamefont {M.}~\bibnamefont {Opitz}},\ }\href@noop {} {\bibfield
  {journal} {\bibinfo  {journal} {J. R. Soc. Interface}\ }\textbf {\bibinfo
  {volume} {11}},\ \bibinfo {pages} {20140172} (\bibinfo {year}
  {2014})}\BibitemShut {NoStop}%
\bibitem [{\citenamefont {Reichenbach}\ \emph
  {et~al.}(2007{\natexlab{a}})\citenamefont {Reichenbach}, \citenamefont
  {Mobilia},\ and\ \citenamefont {Frey}}]{Reichenbach2007}%
  \BibitemOpen
  \bibfield  {author} {\bibinfo {author} {\bibfnamefont {T.}~\bibnamefont
  {Reichenbach}}, \bibinfo {author} {\bibfnamefont {M.}~\bibnamefont
  {Mobilia}}, \ and\ \bibinfo {author} {\bibfnamefont {E.}~\bibnamefont
  {Frey}},\ }\href {\doibase 10.1038/nature06095} {\bibfield  {journal}
  {\bibinfo  {journal} {Nature}\ }\textbf {\bibinfo {volume} {448}},\ \bibinfo
  {pages} {1046} (\bibinfo {year} {2007}{\natexlab{a}})}\BibitemShut {NoStop}%
\bibitem [{\citenamefont {Reichenbach}\ \emph
  {et~al.}(2007{\natexlab{b}})\citenamefont {Reichenbach}, \citenamefont
  {Mobilia},\ and\ \citenamefont {Frey}}]{Reichenbach2007a}%
  \BibitemOpen
  \bibfield  {author} {\bibinfo {author} {\bibfnamefont {T.}~\bibnamefont
  {Reichenbach}}, \bibinfo {author} {\bibfnamefont {M.}~\bibnamefont
  {Mobilia}}, \ and\ \bibinfo {author} {\bibfnamefont {E.}~\bibnamefont
  {Frey}},\ }\href {\doibase 10.1103/PhysRevLett.99.238105} {\bibfield
  {journal} {\bibinfo  {journal} {Phys. Rev. Lett.}\ }\textbf {\bibinfo
  {volume} {99}},\ \bibinfo {pages} {238105} (\bibinfo {year}
  {2007}{\natexlab{b}})}\BibitemShut {NoStop}%
\bibitem [{\citenamefont {A.~Traulsen}\ \emph {et~al.}(2005)\citenamefont
  {A.~Traulsen}, \citenamefont {Claussen},\ and\ \citenamefont
  {Hauert}}]{traulsen-2005-95}%
  \BibitemOpen
  \bibfield  {author} {\bibinfo {author} {\bibfnamefont {A.}~\bibnamefont
  {A.~Traulsen}}, \bibinfo {author} {\bibfnamefont {J.~C.}\ \bibnamefont
  {Claussen}}, \ and\ \bibinfo {author} {\bibfnamefont {C.}~\bibnamefont
  {Hauert}},\ }\href@noop {} {\bibfield  {journal} {\bibinfo  {journal} {Phys.
  Rev. Lett.}\ }\textbf {\bibinfo {volume} {95}},\ \bibinfo {pages} {238701}
  (\bibinfo {year} {2005})}\BibitemShut {NoStop}%
\bibitem [{\citenamefont {Reichenbach}\ \emph {et~al.}(2006)\citenamefont
  {Reichenbach}, \citenamefont {Mobilia},\ and\ \citenamefont
  {Frey}}]{Reichenbach2006}%
  \BibitemOpen
  \bibfield  {author} {\bibinfo {author} {\bibfnamefont {T.}~\bibnamefont
  {Reichenbach}}, \bibinfo {author} {\bibfnamefont {M.}~\bibnamefont
  {Mobilia}}, \ and\ \bibinfo {author} {\bibfnamefont {E.}~\bibnamefont
  {Frey}},\ }\href {\doibase 10.1103/PhysRevE.74.051907} {\bibfield  {journal}
  {\bibinfo  {journal} {Phys. Rev. E}\ }\textbf {\bibinfo {volume} {74}},\
  \bibinfo {pages} {051907} (\bibinfo {year} {2006})}\BibitemShut {NoStop}%
\bibitem [{\citenamefont {Claussen}\ and\ \citenamefont
  {Traulsen}(2008)}]{Claussen2008}%
  \BibitemOpen
  \bibfield  {author} {\bibinfo {author} {\bibfnamefont {J.~C.}\ \bibnamefont
  {Claussen}}\ and\ \bibinfo {author} {\bibfnamefont {A.}~\bibnamefont
  {Traulsen}},\ }\href {\doibase 10.1103/PhysRevLett.100.058104} {\bibfield
  {journal} {\bibinfo  {journal} {Phys. Rev. Lett.}\ }\textbf {\bibinfo
  {volume} {100}},\ \bibinfo {pages} {058104} (\bibinfo {year}
  {2008})}\BibitemShut {NoStop}%
\bibitem [{\citenamefont {Rulands}\ \emph {et~al.}(2011)\citenamefont
  {Rulands}, \citenamefont {Reichenbach},\ and\ \citenamefont
  {Frey}}]{Rulands2011}%
  \BibitemOpen
  \bibfield  {author} {\bibinfo {author} {\bibfnamefont {S.}~\bibnamefont
  {Rulands}}, \bibinfo {author} {\bibfnamefont {T.}~\bibnamefont
  {Reichenbach}}, \ and\ \bibinfo {author} {\bibfnamefont {E.}~\bibnamefont
  {Frey}},\ }\href {\doibase 10.1088/1742-5468/2011/01/L01003} {\bibfield
  {journal} {\bibinfo  {journal} {J. Stat. Mech.}\ }\textbf {\bibinfo {volume}
  {2011}},\ \bibinfo {pages} {L01003} (\bibinfo {year} {2011})}\BibitemShut
  {NoStop}%
\bibitem [{\citenamefont {Traulsen}\ \emph {et~al.}(2012)\citenamefont
  {Traulsen}, \citenamefont {Claussen},\ and\ \citenamefont
  {Hauert}}]{Traulsen2012}%
  \BibitemOpen
  \bibfield  {author} {\bibinfo {author} {\bibfnamefont {A.}~\bibnamefont
  {Traulsen}}, \bibinfo {author} {\bibfnamefont {J.~C.}\ \bibnamefont
  {Claussen}}, \ and\ \bibinfo {author} {\bibfnamefont {C.}~\bibnamefont
  {Hauert}},\ }\href {\doibase 10.1103/PhysRevE.85.041901} {\bibfield
  {journal} {\bibinfo  {journal} {Phys. Rev. E}\ }\textbf {\bibinfo {volume}
  {85}},\ \bibinfo {pages} {041901} (\bibinfo {year} {2012})}\BibitemShut
  {NoStop}%
\bibitem [{\citenamefont {Rulands}\ \emph {et~al.}(2013)\citenamefont
  {Rulands}, \citenamefont {Zielinski},\ and\ \citenamefont
  {Frey}}]{Rulands2013a}%
  \BibitemOpen
  \bibfield  {author} {\bibinfo {author} {\bibfnamefont {S.}~\bibnamefont
  {Rulands}}, \bibinfo {author} {\bibfnamefont {A.}~\bibnamefont {Zielinski}},
  \ and\ \bibinfo {author} {\bibfnamefont {E.}~\bibnamefont {Frey}},\ }\href
  {\doibase 10.1103/PhysRevE.87.052710} {\bibfield  {journal} {\bibinfo
  {journal} {Phys. Rev. E}\ }\textbf {\bibinfo {volume} {87}},\ \bibinfo
  {pages} {052710} (\bibinfo {year} {2013})}\BibitemShut {NoStop}%
\bibitem [{\citenamefont {Dobramysl}\ and\ \citenamefont
  {T\"{a}uber}(2013{\natexlab{a}})}]{Dobramysl2013}%
  \BibitemOpen
  \bibfield  {author} {\bibinfo {author} {\bibfnamefont {U.}~\bibnamefont
  {Dobramysl}}\ and\ \bibinfo {author} {\bibfnamefont {U.~C.}\ \bibnamefont
  {T\"{a}uber}},\ }\href {\doibase 10.1103/PhysRevLett.110.048105} {\bibfield
  {journal} {\bibinfo  {journal} {Phys. Rev. Lett.}\ }\textbf {\bibinfo
  {volume} {110}},\ \bibinfo {pages} {048105} (\bibinfo {year}
  {2013}{\natexlab{a}})}\BibitemShut {NoStop}%
\bibitem [{\citenamefont {Dobramysl}\ and\ \citenamefont
  {T\"{a}uber}(2013{\natexlab{b}})}]{Dobramysl2013a}%
  \BibitemOpen
  \bibfield  {author} {\bibinfo {author} {\bibfnamefont {U.}~\bibnamefont
  {Dobramysl}}\ and\ \bibinfo {author} {\bibfnamefont {U.~C.}\ \bibnamefont
  {T\"{a}uber}},\ }\href {http://stacks.iop.org/1742-5468/2013/i=10/a=P10001}
  {\bibfield  {journal} {\bibinfo  {journal} {J. Stat. Mech.}\ }\textbf
  {\bibinfo {volume} {2013}},\ \bibinfo {pages} {P10001} (\bibinfo {year}
  {2013}{\natexlab{b}})}\BibitemShut {NoStop}%
\bibitem [{\citenamefont {Reichenbach}\ \emph {et~al.}(2008)\citenamefont
  {Reichenbach}, \citenamefont {Mobilia},\ and\ \citenamefont
  {Frey}}]{Reichenbach2008}%
  \BibitemOpen
  \bibfield  {author} {\bibinfo {author} {\bibfnamefont {T.}~\bibnamefont
  {Reichenbach}}, \bibinfo {author} {\bibfnamefont {M.}~\bibnamefont
  {Mobilia}}, \ and\ \bibinfo {author} {\bibfnamefont {E.}~\bibnamefont
  {Frey}},\ }\href
  {http://linkinghub.elsevier.com/retrieve/pii/S0022519308002464
  papers://6bd6bfca-595b-4d6b-94ac-3305a1243fe1/Paper/p184} {\bibfield
  {journal} {\bibinfo  {journal} {J. Theor. Biol}\ }\textbf {\bibinfo {volume}
  {254}},\ \bibinfo {pages} {368} (\bibinfo {year} {2008})}\BibitemShut
  {NoStop}%
\bibitem [{\citenamefont {Knebel}\ \emph {et~al.}(2013)\citenamefont {Knebel},
  \citenamefont {Kr\"{u}ger}, \citenamefont {Weber},\ and\ \citenamefont
  {Frey}}]{Knebel2013}%
  \BibitemOpen
  \bibfield  {author} {\bibinfo {author} {\bibfnamefont {J.}~\bibnamefont
  {Knebel}}, \bibinfo {author} {\bibfnamefont {T.}~\bibnamefont {Kr\"{u}ger}},
  \bibinfo {author} {\bibfnamefont {M.~F.}\ \bibnamefont {Weber}}, \ and\
  \bibinfo {author} {\bibfnamefont {E.}~\bibnamefont {Frey}},\ }\href
  {http://arxiv.org/abs/1303.7116} {\bibfield  {journal} {\bibinfo  {journal}
  {Phys. Rev. Lett.}\ }\textbf {\bibinfo {volume} {110}},\ \bibinfo {pages}
  {168106} (\bibinfo {year} {2013})}\BibitemShut {NoStop}%
\bibitem [{\citenamefont {Szolnoki}\ and\ \citenamefont
  {Szab\'o}(2004)}]{Szabo2004}%
  \BibitemOpen
  \bibfield  {author} {\bibinfo {author} {\bibfnamefont {A.}~\bibnamefont
  {Szolnoki}}\ and\ \bibinfo {author} {\bibfnamefont {G.}~\bibnamefont
  {Szab\'o}},\ }\href {\doibase 10.1103/PhysRevE.70.037102} {\bibfield
  {journal} {\bibinfo  {journal} {Phys. Rev. E}\ }\textbf {\bibinfo {volume}
  {70}},\ \bibinfo {pages} {037102} (\bibinfo {year} {2004})}\BibitemShut
  {NoStop}%
\bibitem [{\citenamefont {Balaban}\ \emph {et~al.}(2004)\citenamefont
  {Balaban}, \citenamefont {Merrin}, \citenamefont {Chait}, \citenamefont
  {Kowalik},\ and\ \citenamefont {Leibler}}]{Balaban2004}%
  \BibitemOpen
  \bibfield  {author} {\bibinfo {author} {\bibfnamefont {N.~Q.}\ \bibnamefont
  {Balaban}}, \bibinfo {author} {\bibfnamefont {J.}~\bibnamefont {Merrin}},
  \bibinfo {author} {\bibfnamefont {R.}~\bibnamefont {Chait}}, \bibinfo
  {author} {\bibfnamefont {L.}~\bibnamefont {Kowalik}}, \ and\ \bibinfo
  {author} {\bibfnamefont {S.}~\bibnamefont {Leibler}},\ }\href {\doibase
  10.1126/science.1099390} {\bibfield  {journal} {\bibinfo  {journal}
  {Science}\ }\textbf {\bibinfo {volume} {305}},\ \bibinfo {pages} {1622}
  (\bibinfo {year} {2004})}\BibitemShut {NoStop}%
\bibitem [{\citenamefont {Gardner}\ \emph {et~al.}(2000)\citenamefont
  {Gardner}, \citenamefont {Cantor},\ and\ \citenamefont
  {Collins}}]{Gardner2000}%
  \BibitemOpen
  \bibfield  {author} {\bibinfo {author} {\bibfnamefont {T.~S.}\ \bibnamefont
  {Gardner}}, \bibinfo {author} {\bibfnamefont {C.~R.}\ \bibnamefont {Cantor}},
  \ and\ \bibinfo {author} {\bibfnamefont {J.~J.}\ \bibnamefont {Collins}},\
  }\href@noop {} {\bibfield  {journal} {\bibinfo  {journal} {Nature}\ }\textbf
  {\bibinfo {volume} {403}},\ \bibinfo {pages} {339} (\bibinfo {year}
  {2000})}\BibitemShut {NoStop}%
\bibitem [{SM()}]{SM}%
  \BibitemOpen
  \href@noop {} {}\bibinfo {note} {See Supplemental Material at \url{http: ...}, which includes Refs.~[31-34], details of the calculations and videos illustrating the population
  dynamics.}\BibitemShut {Stop}%
 \bibitem [{\citenamefont {Szab\'{o}}\ and\ \citenamefont
  {Cz\'{a}r\'{a}n}(2001)}]{Szabo2001}%
  \BibitemOpen
  \bibfield  {author} {\bibinfo {author} {\bibfnamefont {G.}~\bibnamefont
  {Szab\'{o}}}\ and\ \bibinfo {author} {\bibfnamefont {T.}~\bibnamefont
  {Cz\'{a}r\'{a}n}},\ }\href {\doibase 10.1103/PhysRevE.64.042902} {\bibfield
  {journal} {\bibinfo  {journal} {Phys. Rev. E}\ }\textbf {\bibinfo {volume}
  {64}},\ \bibinfo {pages} {042902} (\bibinfo {year} {2001})}\BibitemShut
  {NoStop}%
\bibitem [{\citenamefont {Szab\'{o}}(2005)}]{Szabo2005a}%
  \BibitemOpen
  \bibfield  {author} {\bibinfo {author} {\bibfnamefont {G.}~\bibnamefont
  {Szab\'{o}}},\ }\href
  {http://www.iop.org/EJ/article/0305-4470/38/30/005/a5\_30\_005.pdf
  papers://6bd6bfca-595b-4d6b-94ac-3305a1243fe1/Paper/p186} {\bibfield
  {journal} {\bibinfo  {journal} {J. Phys. A: Math. Gen.}\ }\textbf {\bibinfo
  {volume} {38}},\ \bibinfo {pages} {6689} (\bibinfo {year}
  {2005})}\BibitemShut {NoStop}%
\bibitem [{\citenamefont {Szab\'{o}}\ \emph {et~al.}(2007)\citenamefont
  {Szab\'{o}}, \citenamefont {Cz\'{a}r\'{a}n},\ and\ \citenamefont
  {Szab\'{o}}}]{Szabo2007a}%
  \BibitemOpen
  \bibfield  {author} {\bibinfo {author} {\bibfnamefont {P.}~\bibnamefont
  {Szab\'{o}}}, \bibinfo {author} {\bibfnamefont {T.}~\bibnamefont
  {Cz\'{a}r\'{a}n}}, \ and\ \bibinfo {author} {\bibfnamefont {G.}~\bibnamefont
  {Szab\'{o}}},\ }\href {\doibase 10.1016/j.jtbi.2007.06.022} {\bibfield
  {journal} {\bibinfo  {journal} {J. Theor. Biol.}\ }\textbf {\bibinfo {volume}
  {248}},\ \bibinfo {pages} {736} (\bibinfo {year} {2007})}\BibitemShut
  {NoStop}%
\bibitem [{\citenamefont {Szab\'{o}}\ \emph {et~al.}(2008)\citenamefont
  {Szab\'{o}}, \citenamefont {Szolnoki},\ and\ \citenamefont
  {Borsos}}]{Szabo2008}%
  \BibitemOpen
  \bibfield  {author} {\bibinfo {author} {\bibfnamefont {G.}~\bibnamefont
  {Szab\'{o}}}, \bibinfo {author} {\bibfnamefont {A.}~\bibnamefont {Szolnoki}},
  \ and\ \bibinfo {author} {\bibfnamefont {I.}~\bibnamefont {Borsos}},\ }\href
  {\doibase 10.1103/PhysRevE.77.041919} {\bibfield  {journal} {\bibinfo
  {journal} {Phys. Rev. E}\ }\textbf {\bibinfo {volume} {77}},\ \bibinfo
  {pages} {041919} (\bibinfo {year} {2008})}\BibitemShut {NoStop}%
\bibitem [{\citenamefont {Lotka}(1920)}]{lotka1920}%
  \BibitemOpen
  \bibfield  {author} {\bibinfo {author} {\bibfnamefont {A.~J.}\ \bibnamefont
  {Lotka}},\ }\href {\doibase 10.1021/ja01453a010} {\bibfield  {journal}
  {\bibinfo  {journal} {J. Am. Chem. Soc.}\ }\textbf {\bibinfo {volume} {42}},\
  \bibinfo {pages} {1595} (\bibinfo {year} {1920})}\BibitemShut {NoStop}%
\bibitem [{\citenamefont {Volterra}(1926)}]{volterra-1926-31}%
  \BibitemOpen
  \bibfield  {author} {\bibinfo {author} {\bibfnamefont {V.}~\bibnamefont
  {Volterra}},\ }\href@noop {} {\bibfield  {journal} {\bibinfo  {journal} {Mem.
  Accad. Lincei}\ }\textbf {\bibinfo {volume} {2}},\ \bibinfo {pages} {31}
  (\bibinfo {year} {1926})}\BibitemShut {NoStop}%
\bibitem [{\citenamefont {Lehe~R.}(2012)}]{Lehe2012}%
  \BibitemOpen
  \bibfield  {author} {\bibinfo {author} {\bibfnamefont {P.~L.}\ \bibnamefont
  {Lehe~R.}, \bibfnamefont {Hallatschek~O.}},\ }\href@noop {} {\bibfield
  {journal} {\bibinfo  {journal} {PLoS Comput. Biol.}\ }\textbf {\bibinfo
  {volume} {8}},\ \bibinfo {pages} {e1002447} (\bibinfo {year}
  {2012})}\BibitemShut {NoStop}%
\bibitem [{\citenamefont {Korolev}\ \emph {et~al.}(2012)\citenamefont
  {Korolev}, \citenamefont {M{\"u}ller}, \citenamefont {Karahan}, \citenamefont
  {Murray}, \citenamefont {Hallatschek},\ and\ \citenamefont
  {Nelson}}]{Korolev2012}%
  \BibitemOpen
  \bibfield  {author} {\bibinfo {author} {\bibfnamefont {K.~S.}\ \bibnamefont
  {Korolev}}, \bibinfo {author} {\bibfnamefont {M.~J.~I.}\ \bibnamefont
  {M{\"u}ller}}, \bibinfo {author} {\bibfnamefont {N.}~\bibnamefont {Karahan}},
  \bibinfo {author} {\bibfnamefont {A.~W.}\ \bibnamefont {Murray}}, \bibinfo
  {author} {\bibfnamefont {O.}~\bibnamefont {Hallatschek}}, \ and\ \bibinfo
  {author} {\bibfnamefont {D.~R.}\ \bibnamefont {Nelson}},\ }\href
  {http://stacks.iop.org/1478-3975/9/i=2/a=026008} {\bibfield  {journal}
  {\bibinfo  {journal} {Phys. Biol.}\ }\textbf {\bibinfo {volume} {9}},\
  \bibinfo {pages} {026008} (\bibinfo {year} {2012})}\BibitemShut {NoStop}%
\bibitem [{\citenamefont {May}\ and\ \citenamefont {Leonard}(1975)}]{May1975}%
  \BibitemOpen
  \bibfield  {author} {\bibinfo {author} {\bibfnamefont {R.~M.}\ \bibnamefont
  {May}}\ and\ \bibinfo {author} {\bibfnamefont {W.~J.}\ \bibnamefont
  {Leonard}},\ }\href {http://link.aip.org/link/?SMJMAP/29/243/1
  papers://6bd6bfca-595b-4d6b-94ac-3305a1243fe1/Paper/p197} {\bibfield
  {journal} {\bibinfo  {journal} {SIAM J. Appl. Math.}\ }\textbf {\bibinfo
  {volume} {29}},\ \bibinfo {pages} {243} (\bibinfo {year} {1975})}\BibitemShut
  {NoStop}%
\bibitem [{Note1()}]{Note1}%
  \BibitemOpen
  \bibinfo {note} {This can easily be generalized to asymmetric competition
  between phenotypes by considering phenotype dependent competition
  rates.}\BibitemShut {Stop}%
\bibitem [{\citenamefont {Frey}(2010)}]{frey2010}%
  \BibitemOpen
  \bibfield  {author} {\bibinfo {author} {\bibfnamefont {E.}~\bibnamefont
  {Frey}},\ }\href {\doibase 10.1016/j.physa.2010.02.047} {\bibfield  {journal}
  {\bibinfo  {journal} {Physica A}\ }\textbf {\bibinfo {volume} {389}},\
  \bibinfo {pages} {4265} (\bibinfo {year} {2010})}\BibitemShut {NoStop}%
\bibitem [{\citenamefont {Reichenbach}\ and\ \citenamefont
  {Frey}(2008)}]{Reichenbach2008b}%
  \BibitemOpen
  \bibfield  {author} {\bibinfo {author} {\bibfnamefont {T.}~\bibnamefont
  {Reichenbach}}\ and\ \bibinfo {author} {\bibfnamefont {E.}~\bibnamefont
  {Frey}},\ }\href {\doibase 10.1103/PhysRevLett.101.058102} {\bibfield
  {journal} {\bibinfo  {journal} {Phys. Rev. Lett.}\ }\textbf {\bibinfo
  {volume} {101}},\ \bibinfo {pages} {058102} (\bibinfo {year}
  {2008})}\BibitemShut {NoStop}%
\bibitem [{\citenamefont {Szczesny}\ \emph {et~al.}(2013)\citenamefont
  {Szczesny}, \citenamefont {Mobilia},\ and\ \citenamefont
  {Rucklidge}}]{Szczesny2013}%
  \BibitemOpen
  \bibfield  {author} {\bibinfo {author} {\bibfnamefont {B.}~\bibnamefont
  {Szczesny}}, \bibinfo {author} {\bibfnamefont {M.}~\bibnamefont {Mobilia}}, \
  and\ \bibinfo {author} {\bibfnamefont {A.~M.}\ \bibnamefont {Rucklidge}},\
  }\href {\doibase 10.1209/0295-5075/102/28012} {\bibfield  {journal} {\bibinfo
   {journal} {Europhys. Lett.}\ }\textbf {\bibinfo {volume} {102}},\ \bibinfo
  {pages} {28012} (\bibinfo {year} {2013})}\BibitemShut {NoStop}%
\bibitem [{\citenamefont {Sheftel}\ \emph {et~al.}(2013)\citenamefont
  {Sheftel}, \citenamefont {Shoval}, \citenamefont {Mayo},\ and\ \citenamefont
  {Alon}}]{Sheftel2013}%
  \BibitemOpen
  \bibfield  {author} {\bibinfo {author} {\bibfnamefont {H.}~\bibnamefont
  {Sheftel}}, \bibinfo {author} {\bibfnamefont {O.}~\bibnamefont {Shoval}},
  \bibinfo {author} {\bibfnamefont {A.}~\bibnamefont {Mayo}}, \ and\ \bibinfo
  {author} {\bibfnamefont {U.}~\bibnamefont {Alon}},\ }\href {\doibase
  10.1002/ece3.528} {\bibfield  {journal} {\bibinfo  {journal} {Ecol. Evol.}\
  }\textbf {\bibinfo {volume} {3}},\ \bibinfo {pages} {1471} (\bibinfo {year}
  {2013})}\BibitemShut {NoStop}%
\bibitem [{Note2()}]{Note2}%
  \BibitemOpen
  \bibinfo {note} {Our simulations show that the three surviving genotypes in
  the metastable regime contribute with equal probability to the asymptotic
  states. For the reason of numerical efficiency we therefore computed
  $P_\infty (\protect \mathaccentV {vec}17E{\pi })$ at times corresponding to
  the metastable regime.}\BibitemShut {Stop}%
\bibitem [{Note3()}]{Note3}%
  \BibitemOpen
  \bibinfo {note} {For symmetry reasons all three of these marginal
  distributions are identical.}\BibitemShut {Stop}%
\bibitem [{\citenamefont {Hofbauer}\ and\ \citenamefont
  {Sigmund}(1998)}]{Hofbauer1998}%
  \BibitemOpen
  \bibfield  {author} {\bibinfo {author} {\bibfnamefont {J.}~\bibnamefont
  {Hofbauer}}\ and\ \bibinfo {author} {\bibfnamefont {K.}~\bibnamefont
  {Sigmund}},\ }\href@noop {} {\emph {\bibinfo {title} {{Evolutionary Games and
  Population Dynamics}}}},\ \bibinfo {edition} {1st}\ ed.\ (\bibinfo
  {publisher} {Cambridge University Press},\ \bibinfo {address} {Cambridge},\
  \bibinfo {year} {1998})\BibitemShut {NoStop}%
\bibitem [{\citenamefont {Parker}\ and\ \citenamefont
  {Kamenev}(2009)}]{Parker2009}%
  \BibitemOpen
  \bibfield  {author} {\bibinfo {author} {\bibfnamefont {M.}~\bibnamefont
  {Parker}}\ and\ \bibinfo {author} {\bibfnamefont {A.}~\bibnamefont
  {Kamenev}},\ }\href {\doibase 10.1103/PhysRevE.80.021129} {\bibfield
  {journal} {\bibinfo  {journal} {Phys. Rev. E}\ }\textbf {\bibinfo {volume}
  {80}},\ \bibinfo {pages} {021129} (\bibinfo {year} {2009})}\BibitemShut
  {NoStop}%
\bibitem [{\citenamefont {Dobrinevski}\ and\ \citenamefont
  {Frey}(2012)}]{Dobrinevski2012}%
  \BibitemOpen
  \bibfield  {author} {\bibinfo {author} {\bibfnamefont {A.}~\bibnamefont
  {Dobrinevski}}\ and\ \bibinfo {author} {\bibfnamefont {E.}~\bibnamefont
  {Frey}},\ }\href {\doibase 10.1103/PhysRevE.85.051903} {\bibfield  {journal}
  {\bibinfo  {journal} {Phys. Rev. E}\ }\textbf {\bibinfo {volume} {85}},\
  \bibinfo {pages} {051903} (\bibinfo {year} {2012})}\BibitemShut {NoStop}%
\bibitem [{\citenamefont {Dobrinevski}\ \emph {et~al.}(2014)\citenamefont
  {Dobrinevski}, \citenamefont {Alava}, \citenamefont {Reichenbach},\ and\
  \citenamefont {Frey}}]{Drobrinevski2014}%
  \BibitemOpen
  \bibfield  {author} {\bibinfo {author} {\bibfnamefont {A.}~\bibnamefont
  {Dobrinevski}}, \bibinfo {author} {\bibfnamefont {M.}~\bibnamefont {Alava}},
  \bibinfo {author} {\bibfnamefont {T.}~\bibnamefont {Reichenbach}}, \ and\
  \bibinfo {author} {\bibfnamefont {E.}~\bibnamefont {Frey}},\ }\href {\doibase
  10.1103/PhysRevE.89.012721} {\bibfield  {journal} {\bibinfo  {journal} {Phys.
  Rev. E}\ }\textbf {\bibinfo {volume} {89}},\ \bibinfo {pages} {012721}
  (\bibinfo {year} {2014})}\BibitemShut {NoStop}%
\end{thebibliography}
%

\cleardoublepage

\pagestyle{empty}


\section*{Supplementary material}

# Supplementary Material to Specialization and Bet-Hedging in Heterogeneous Populations

Steffen Rulands, David Jahn, and Erwin Frey

*Department of Physics, Arnold-Sommerfeld-Center for Theoretical Physics and Center for NanoScience,*
*Ludwig-Maximilians-Universität München, Theresienstrasse 37, D-80333 Munich, Germany*

In the Supplementary Material we provide calculations and numerical results supporting the arguments presented in the main text.

## SCALING OF THE CROSSOVER TIME $t_1$ WITH THE SYSTEM SIZE

The time $t_1$ marks the crossover from neutral evolution to selection driven evolution. To determine how it scales with the system size $N$ we try - motivated by the phase portraits of the classical LV and ML models - the following scaling ansatz for the average number of distinct genotypes:

$$H_{\mathrm{LV}}(t, N) = Nt^{-1}h_{\mathrm{LV}}\left(\frac{t}{N}\right), \qquad \text{and} \qquad H_{\mathrm{ML}}(t, N) = Nt^{-1}h_{\mathrm{ML}}\left(\frac{t}{\ln N}\right), \tag{1}$$

where $h_{\mathrm{LV}}$ and $h_{\mathrm{ML}}$ are scaling functions of the heterogeneous LV and the ML model, respectively. As can be inferred from Fig. 1, this scaling ansatz works very well in the relevant time window, and, therefore, the crossover time $t_1$ scales as $\langle t_1 \rangle \propto N$ and $\langle t_1 \rangle \propto \ln N$ for the models of LV and ML type, respectively.

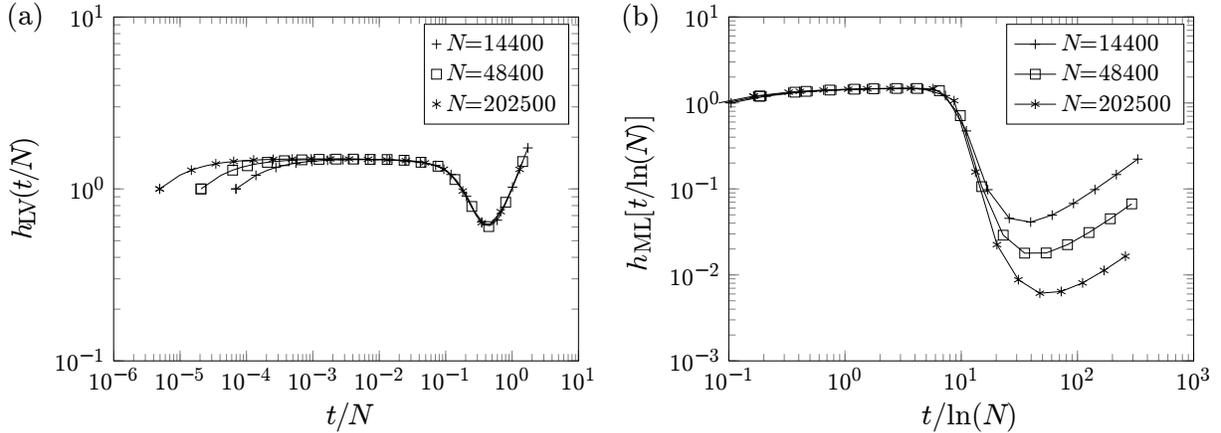

FIG. 1. *Scaling functions h* (a) for the Lotka-Volterra and (b) for the May-Leonard model. The simulations were performed for well-mixed systems, and different system sizes $N$ as indicated in the graph.

## A GEOMETRIC INTERPRETATION OF THE NET TRANSITION RATES

In this section we give a geometric interpretation of the net transition rates $W_{\alpha\beta}$ on the simplex $\Delta^2$, which signify the rates by which genotype $\alpha$ defeats another genotype $\beta$. The net transition rates $W_{\alpha\beta}$ are bilinear forms $W(\vec{p}_\alpha, \vec{p}_\beta) := W_{\alpha\beta}$ in the vectors $\vec{p}_\alpha$ and $\vec{p}_\beta$ characterizing the genotypes $\alpha$ and $\beta$, respectively. For a given genotype $\alpha$, $W(\vec{p}_\alpha, \vec{x}) = s$ (with $s$ some constant and $\vec{x} \in \mathbb{R}^3$) defines a plane whose intersection with the simplex ($\vec{p}_\beta \in \Delta^2$) is a line. Hence, the isolines $W(\vec{p}_\alpha, \vec{p}_\beta) = s$ are those genotypes $\beta$ which compete with $\alpha$ at the same rate $s$. We signify the corresponding set of *parallel* lines by $\{I_\alpha^s\}$. A particular representative of this set of lines are those where the competing genotypes are selectively neutral: $W(\vec{p}_\alpha, \vec{p}_\beta) = 0$. Since $W_{\alpha\alpha} = 0$ and $W_{\alpha B} = 0$, the corresponding line $I_\alpha^0$ runs through $\vec{p}_\alpha$ and the center of the simplex, $\vec{p}_B = (\frac{1}{3}, \frac{1}{3}, \frac{1}{3})$ [Fig. 2(a)]. This neutral isoline $I_\alpha^0$ divides the simplex into two regimes, $\Delta^+$ and $\Delta^-$, to the right (+) and left (−) with respect to the direction pointing from $\alpha$ to the center $B$ of the simplex $\Delta^2$. As $W$ is linear in both of its arguments, in particular in $\vec{p}_\beta$, the net transition rate $W_{\alpha\beta} = s$ is a monotonically

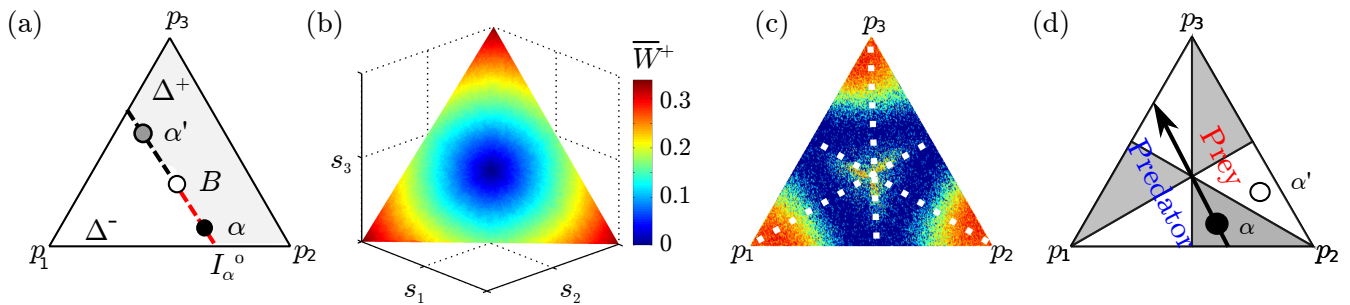

FIG. 2. *Geometric interpretation for the net transition rates $W_{\alpha\beta}$*. (a) Genotypes $\alpha$ (closed circle) correspond to points on the two-dimensional simplex $\triangle^2$. The dashed line, denoted by $I_\alpha^0$, indicates genotypes $\beta$ which are neutral with respect to this genotype $\alpha$: $W_{\alpha\beta} = 0$. It can be constructed by drawing a line through $\alpha$ and the strict bet-hedging genotype $B = (1/3, 1/3, 1/3)$ (open circle). Genotypes lying on the red portion of this line outcompete the same genotypes as $\alpha$ (shaded area, $\triangle^+$) and, at the same time, are outcompeted by the same genotypes as $\alpha$ (white area, $\triangle^-$). (b) Selective power $\overline{W}^+$ for genotypes on the simplex $\triangle^2$, where red denotes large values of $\overline{W}_\alpha^+$, and blue signifies small values of $\overline{W}_\alpha^+$. We find that $\overline{W}_\alpha^+$ increases linearly with the distance from the center and is highest for specialists. $\overline{W}_\alpha^+$ is constant along circles around the center. (c) To indicate strategic associations in the regime corresponding to small values of $D$, we reproduce Fig. 3(a), top left, from the main text with isolines for the three specialists. Color denotes the probability density of asymptotic genotypes, such that red signifies successful genotypes and blue signifies unsuccessful genotypes. One can see that, for each specialists, there is a surviving bet-hedging genotype on the same isoline. (d) Solid lines indicate genotypes which are neutral with respect to one of the specialist genotypes. Genotypes in shaded regions (e.g. $\alpha$, closed circle) outcompete a larger number of genotypes than genotypes in the white regions (e.g. $\alpha'$, open circle). The arrow indicates the isoline of genotype $\alpha$ (closed circle). All genotypes right to this line are prey and all genotypes left to this line are predator with respect to $\alpha$.

increasing/decreasing function of the distance $d$ to $I_\alpha^0$ in $\triangle^{+/-}$; recall that the isolines $I_\alpha^s$ are parallel to $I_\alpha^0$. Note also that the sign of the slope of $s(d)$ is simply defined by the rules of cyclic dominance. Hence for a genotype $\alpha'$, on the opposite side of $B$, the roles of $\triangle^+$ and $\triangle^-$ are interchanged.

Taken together this implies that the best response (largest value of $W_{\alpha\beta}$) to $\alpha$ is a genotype $\beta$ with the largest distance from $I_\alpha^0$, *i.e.* a genotype on the boundary of the simplex in regime $\triangle^-$. For the example in Fig. 2(a) the best response to $\alpha$ is the specialist $\vec{p}_1$, and the best response to $\alpha'$ is $\vec{p}_3$. In short, the best response to any genotype $\vec{p}_\alpha \in \triangle^2$ is a genotype on the boundary of the simplex $\triangle^2$. If the line $I_\alpha^0$ is not parallel to one of the borders, the best response against the genotype $\alpha$ is unique and given by one of the three specialist genotypes. In conclusion, the best and worst responses to almost all genotypes in $\triangle^2$ are specialists.

From the monotonicity of $W$ we can draw some further conclusions: Since $W_{\alpha B} = 0$ for any genotype on $\alpha \in \triangle^2$, the slope of $s(d)$ must decrease with $\alpha$ approaching the center $B$ of the simplex. Hence the best response to a genotype $\alpha$ is weakest for bet-hedgers (those close to the center of the simplex) and strongest for specialists (those closest to the boundary of the simplex). This implies that with increasing distance from the center of the simplex the value of the expected net invasion rate for genotype $\alpha$ ('selective power') [1],

$$\overline{W}_\alpha^+ := \sum_{\beta \in \triangle^+} W_{\alpha\beta}\,,\qquad(2)$$

is a monotonically increasing function [Fig. 2(b)]. Figure 2(b) also shows that $\overline{W}_\alpha^+$ exhibits a rotational symmetry, *i.e.* all genotype $\alpha$ with the same distance from the center $B$ have identical selective power. This symmetry is simply due to fact that we have cyclic dominance with all equal rates. Note that the above arguments do not rely on a specific choice of the interaction matrix $\mathbf{A}$. Rather, we only assumed that net interactions of genotypes with themselves and with the bet-hedging genotype is zero. Then, since specialists define the outer hull of the set of genotypes, the monotonicity of $W$ and the following line of argumentation should hold for a broad class of models satisfying these conditions.

In the following we will employ the above geometric interpretation of the net reaction rate in order to understand the struggle for survival in spatially extended systems.

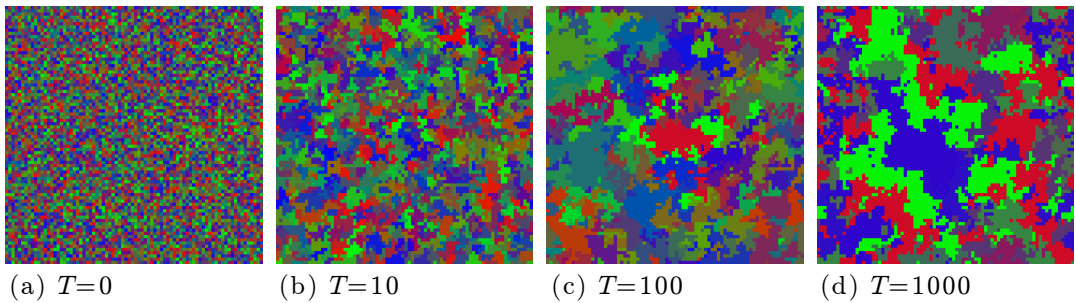

(a) $T=0$     (b) $T=10$     (c) $T=100$     (d) $T=1000$

FIG. 3. *Exemplary sequence of spatial patterns for small values of the diffusion constant.* Different colors signify the probability to be in either of three phenotypic states: red, green or blue denotes a high probability to be in the phenotypic states $s_1$, $s_2$ or $s_3$, respectively, while grey signifies the generalist genotype $B$. Mutually neutral genotypes are represented by different saturations of the same shade of red, blue and green. We observe the formation of clusters of mutually neutral genotypes [(b) and (c)] and ultimately the survival of specialists [(d)]. Parameters were $L = 100$ and $D = 0$.

## SURVIVAL OF SPECIALISTS FOR SMALL MOBILITIES

How can we understand the dominance of specialists for very small values of the diffusion constant $D$? In a spatially extended system the key quantity to consider is the magnitude of $W_{\alpha\beta}$. It determines the speed and direction of propagation of domain borders between genotypes, and thereby the outcome of pairwise competition.

Starting from an initial state with a high degree of genetic diversity the system first coarsens into spatial domains containing pairwise selectively neutral genotypes [Fig. 3]. Consider now a spatial cluster comprised of such selectively equivalent genotypes $\alpha$ [e.g. all genotypes on the dashed red line in Fig. 2(a)]. As we have learned in the previous section, the closer a genotype $\alpha$ is to the boundary of the simplex, the larger is its *expected* rate of being invaded ($\overline{W_\alpha^-}$) or to invade ($\overline{W_\alpha^+}$) domains of any other genotype not lying on the isoline $I_\alpha^0$. Hence, specialists within such a cluster have a higher net reproduction rate in boundary regions adjoining domains of their respective prey ($\Delta^+$ region), and, therefore, are able to invade such regions fast. Pictorially speaking, specialists are "good in offense". In contrast, generalists have a lower overall rate of being invaded, and, therefore, they will dominate in boundary regions facing their respective predators ($\Delta^-$ region); they are "good defenders". This leads to an internal organisation of the selectively neutral clusters, where generalists and specialist form strategic alliances [2–5]. Those profiting the most from this alliance are the specialists since it enables them to invade new territory fast. Taken together, this leads to a "first come first served" principle in the sense that those with the fastest expected invasion rate $\overline{W_\alpha^+}$ will dominate the population in the long run.

As a final aside we note that while specialists are the most dominant genotype the population for large times they are actually not the only surviving genotypes. In addition to each specialist genotype there is an associated bet-hedging genotype located on the same isoline $I_\alpha^0$ as the specialist, cf. the yellow areas in Fig. 2(c) with the isoline indicated as a white dashed line. Recall that the initial coarsening process leads to spatial domains containing pairwise neutral genotypes and, geometrically, those are located along the neutral isoline $I_\alpha^0$ on the simplex. For specificity let's pick the specialist $\vec{p}_1$. Then, the associated bet-hedger is given by $\vec{p}_{bet-hedge} \approx (\frac{1}{3} - 2\epsilon, \frac{1}{3} + \epsilon, \frac{1}{3} + \epsilon)$ with $\epsilon$ some small positive number: Together they form a strategic alliance where $\vec{p}_{bet-hedge}$ protects $\vec{p}_1$ from its predator $\vec{p}_3$ due to the enhanced probability to be in phenotypic state $s_2$.

## BIASED BET HEDGING FOR INTERMEDIATE MOBILITIES

The regime with intermediate values of the diffusion constant $D$ is characterised by the survival of biased bet-hedgers as mentioned in the main text. In order to survive they need to balance invasion speed with robustness against hostile invasion. Hence, the most successful genotypes can be neither full specialists nor full bet-hedging genotypes. Surprisingly, our numerical simulations show that their degree of bet-hedging is not only determined by the radial distance from the center of the simplex. Actually, we find that the survival probability is highest in three of the six triangles defined by the perpendicular bisectors of the simplex [shaded regions in Fig. 2(d)]. In the following we will explain (on the basis of the geometric interpretation of the net rates $W_{\alpha\beta}$) why these particular biased bed-hedging genotypes are evolutionary most successful.

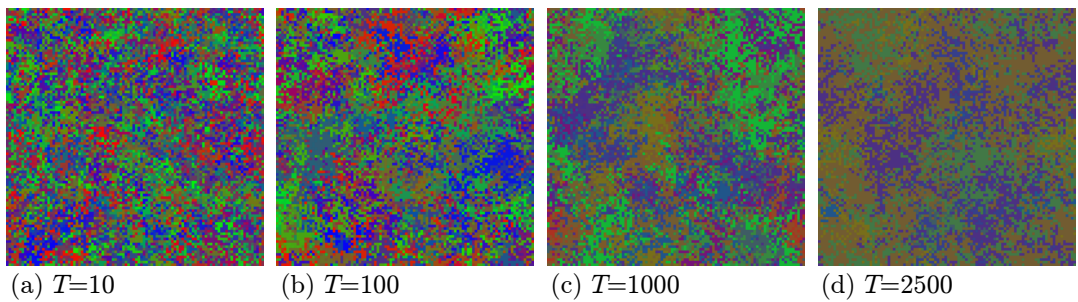

(a) $T=10$  (b) $T=100$  (c) $T=1000$  (d) $T=2500$

FIG. 4. *Exemplary sequence of spatial patterns for intermediate values of the diffusion constant.* Different colors signify the probability to be in either of three phenotypic states: red, green or blue denotes a high probability to be in the phenotypic states $s_1$, $s_2$ or $s_3$, respectively, while grey signifies the generalist genotype $B$. Mutually neutral genotypes are represented by different saturations of the same shade of red, blue and green. While we still observe clusters of mutually similar genotypes[(a)-(c)], domain boundaries are fuzzy and genotypes are frequently subject to random invasion. Parameters were $L = 100$ and $D = 10^{-4}$.

Similar as for low values of $D$ there is an initial coarsening process, but with the domain boundaries less well defined [Fig. 4]. Then, for a given phenotype $\alpha$, its evolutionary success is given by how well it is able to invade territories of its prey while still being able to successfully defend against its predators.

We first observe that genotypes in the gray triangles have more preys than predators, and, therefore, outcompete more genotypes than they are outcompeted by: Consider a genotype $\alpha$ in one of the gray triangles [Fig. 2(d)]. As we have learned from the previous discussions, the prey and predator of a given genotype are located in the areas $\Delta^+$ and $\Delta^-$, respectively. Basic geometry tells us that $1 < Z_\alpha^+/Z_\alpha^- < 5/4$, where $Z^{+/-}$ are the number of genotypes in areas $\Delta^{+/-}$, respectively, *i.e.* the number of prey and predator of $\alpha$. In contrast, for genotypes $\alpha'$ in the white triangles we find $4/5 < Z_{\alpha'}^+/Z_{\alpha'}^- < 1$.

Second, we recall that due to cyclic symmetry the expected overall invasion rate $\overline{W}_\alpha^+$ is constant on circles surrounding the center of the simplex. For a pair of genotypes, $\alpha$ and $\alpha'$, this implies the inequality for the expected invasion rate against a particular prey,

$$\frac{1}{Z_\alpha^+}\sum_{\beta\in\Delta_\alpha^+} W_{\alpha\beta} < \frac{1}{Z_{\alpha'}^+}\sum_{\beta\in\Delta_{\alpha'}^+} W_{\alpha'\beta}\,. \tag{3}$$

Therefore, as the areas of all triangles are the same, the *average* invasion rate for a randomly picked individual in a gray area is lower than in the white areas. We conclude that the surviving genotypes achieve robustness against random invaders by outcompeting a larger set of genotypes in cost of a lower average invasion rate.

As a final aside, we observe that the most successful genotypes obey a hierarchy in the components of the probability distributions $\vec{p}_\alpha$: They have a relatively high probability to be in one phenotypic state, the genotype's bias. The ensuing phenotype with the highest rate of invasion takes the second largest value. Last, the component which is dominated by the genotype's bias has the lowest probability. Mathematically speaking, the components obey $p_\alpha^1 > p_\alpha^3 > p_\alpha^2$, or cyclically. This hierarchy of phenotypic states ensures that domains are less susceptible to invasion by the most aggressive genotypes.

## THE HETEROGENEOUS MAY-LEONARD MODEL

To investigate the evolution of phenotypic heterogeneity in indirect competition we investigate situations, where competition is mediated by the limited availability of resources. In this case, for the ML model, we find a remarkably different behavior. There, independent of the value of the diffusion constant $D$, the population is asymptotically dominated by specialists (Fig. 5). This can be understood as follows: Self-interactions between genetically identical individuals are potentially disadvantageous, as they may lead to the creation of empty sites which in turn may then be colonized by individuals of a different genotype. Since self-interactions are impossible for specialists ($w_{ii} = 1 - \mathbf{p}_i{}^2 = 0$), they are typically better off than their bet-hedging competitors. This advantage is most pronounced at low mobilities where the formation of stable spatial structures is inhibited by reactions between identical genotypes, and thereby promotes the breakup of compact spatial domains.

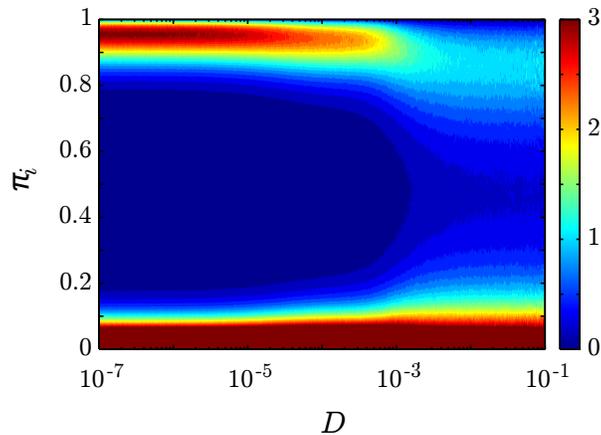

FIG. 5. Marginal probability distribution to be in any of the components of $\vec{\pi}$ for the heterogeneous May-Leonard model. We find that for any value of $D$ specialists dominate the population for large times ($L = 80$). The histograms for each diffusion constant were calculated over $10^5$ trajectories.

For cyclic competition between three specialists the mean field dynamics is described by the May-Leonard equations. In this section, we study the nonlinear dynamics of the heterogeneous version of the May-Leonard model and argue why specialists dominate the population in the long term. For large particle numbers, and for well-mixed systems the dynamics of the heterogeneous May-Leonard model is aptly described by $G$ coupled differential equations for the concentrations of the genotypes $\alpha$,

$$\dot{x}_\alpha = x_\alpha \left[ \mu \left( 1 - \sum_{\beta=1}^{G} x_\beta \right) - \vec{p}_\alpha^{\mathrm{T}} \mathbf{A}^{\mathrm{T}} \sum_{\beta=1}^{G} x_\beta \vec{p}_\beta \right], \quad \alpha \in 1, \ldots, G. \tag{4}$$

The first term on the right hand side does not depend on the specific genotype $\vec{p}_\alpha$ and it gives the rate at which empty sites are populated. The second term is the rate at which individuals are replaced by empty sites. In the following we will show that these equations show qualitatively similar behavior as the May-Leonard equations which leads to the dominance of specialists in the heterogeneous model. To study the nonlinear dynamics in more detail we rewrite Eq. (4) in terms of the $k^{\mathrm{th}}$ "moments" of $\vec{p}_\alpha$:

$$\sigma_{2k} \equiv \sum_\alpha (\vec{p}_\alpha)^{2k} x_\alpha, \tag{5}$$

$$\vec{\sigma}_{2k+1} \equiv \sum_\alpha (\vec{p}_\alpha)^{2k+1} x_\alpha. \tag{6}$$

With this definition, $\sigma_0 \equiv \sum_\alpha x_\alpha$ gives the total concentration of individuals. The first moment gives the mean genotype, $\vec{\sigma}_1 \equiv \sum_\alpha \vec{p}_\alpha x_\alpha$, and the second moment, $\sigma_2 \equiv \sum_\alpha x_\alpha \vec{p}_\alpha \cdot \vec{p}_\alpha$, can be interpreted as the degree of specialization in the population. To obtain the time evolution of the $k^{\mathrm{th}}$ moment we multiply Eqs. (4) by $(\vec{p}_\alpha)^k$ and sum over all genotypes $\alpha$. We obtain for the time evolution of the moments

$$\dot{\sigma}_{2k} = \mu \sigma_{2k}(1 - \sigma_0) - \vec{\sigma}_{2k+1}^{\mathrm{T}} \mathbf{A}^{\mathrm{T}} \vec{\sigma}_1, \tag{7}$$

$$\dot{\vec{\sigma}}_{2k+1} = \mu \vec{\sigma}_{2k+1}(1 - \sigma_0) - \sigma_{2k+2} \mathbf{A}^{\mathrm{T}} \vec{\sigma}_1. \tag{8}$$

For simplicity, we restrict the following analysis to cyclic competition between $M = 3$ phenotypes:

$$\mathbf{A} = \begin{pmatrix} 0 & 1 & 0 \\ 0 & 0 & 1 \\ 1 & 0 & 0 \end{pmatrix}. \tag{9}$$

One immediately sees that in this case $\sigma_0 = 3/4, \vec{\sigma}_{2k-1} = (1/4, \ldots, 1/4), \sigma_{2k} = 1/4$ for $k > 0$, is a (reactive) fixed point of the dynamics. In fact, the initial conditions studied in the main text correspond to this fixed point. Furthermore, the model exhibits $G$ absorbing states corresponding to the extinction of all but one genotype.

We are now interested in the evolutionary dynamics in the vicinity of the reactive fixed point. To this end, we neglect moments of order three and higher and linearize Eqs. (7-8) around this fixed point. The eigenvalues of the

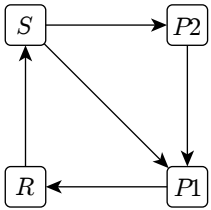

FIG. 6. Interaction network for the heterogeneous, asymmetric four-species model.

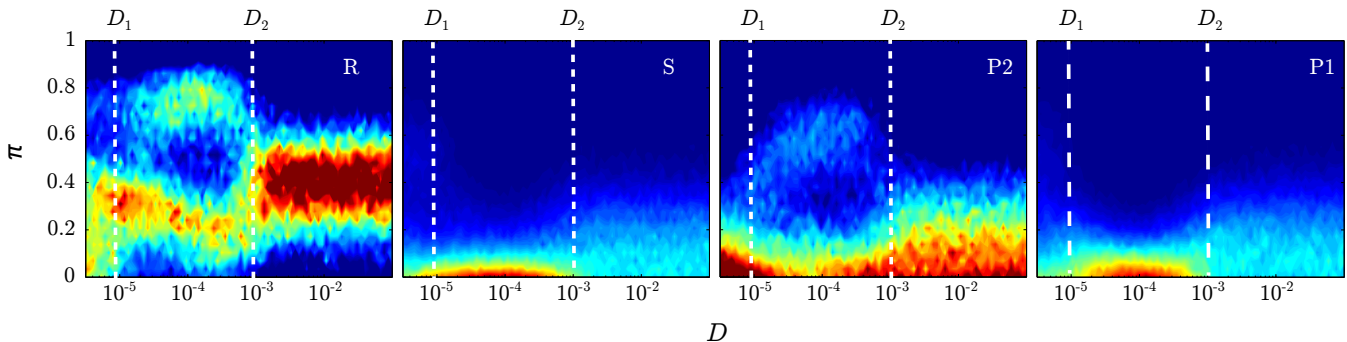

FIG. 7. Marginal probability distribution to be in any of the components of $\vec{\pi}$ for the asymmetric four species model with direct competition. We identify two threshold values separating distinct outcomes of the evolutionary dynamics ($L = 80$). The histograms for each diffusion constant were calculated over $10^5$ trajectories.

Jacobian then determine the exponential dynamics in this region. The linear stability analysis reveals a stable and an unstable eigendirection. In addition, there is a pair of complex conjugate eigenvalues with positive real parts indicating an oscillatory escape out of the reactive fixed point.

Interestingly, this behaviour reminds us of the homogeneous May-Leonard model, where the dynamics spirals outward of an unstable fixed point, approaching states, where only a single species survives. Indeed, numerical simulations of the full dynamics in a well-mixed system show a strikingly similar behaviour for the average genotype $\vec{\sigma}_1$: $\vec{\sigma}_1$ spirals outward of the reactive fixed point approaching the boundary of the simplex and the absorbing states. For the evolution of phenotypic heterogeneity this means that specialists dominate the population at large times. Indeed, in striking contrast to the Lotka-Volterra dynamics, we find the survival of specialists for any value of the diffusion constant $D$, see Fig. 5.

As an aside, we note that even though the heterogeneous May-Leonard model converges to three cyclically competing pure species the asymptotic dynamics is nevertheless slightly different to the homogeneous May-Leonard model. As noted above, the time scale of net competition increases monotonically with an increasing degree of specialization. Furthermore, in a finite heterogeneous system, the surviving genotypes are rarely ideal specialists. Rather, the asymptotic genotypes follow distribution as shown in Fig. 3(a) in the main text. Consequently, the time scale of interaction between the remaining three genotypes is lower in the heterogeneous model, which translates, for a given diffusion constant, to an increase in the length scale of spatio-temporal patterns. Therefore, the previously observed threshold in the diffusion constant [6, 7] is shifted towards lower values of the diffusion constant. In other words, the range of parameters allowing for the coexistence of species is reduced, such that in this sense phenotypic heterogeneity ultimately decreases genetic diversity for very large times.

## THE HETEROGENEOUS ASYMMETRIC FOUR SPECIES MODEL

As an example for more complex interactions we consider a heterogeneous model comprising asymmetric interactions between four phenotypic states P1, P2, S and R [Fig. 6]. The corresponding interaction matrix is

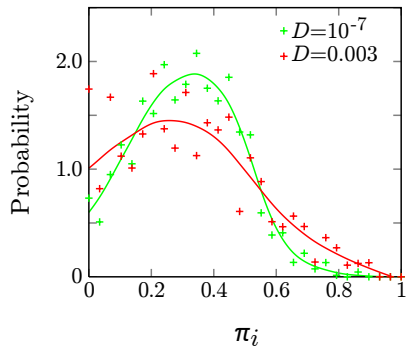

FIG. 8. Marginal probability distribution to be in any of the components of $\boldsymbol{\pi}$ in the LV model in one spatial dimension and for different values of the diffusion constant [Histogram (+) and spline interpolation (solid lines)]. In contrast to the two-dimensional system, we find that asymptotically the population is dominated by bet-hedgers in both cases.

$$\mathbf{A} = \begin{pmatrix} 0 & 1 & 0 & 0 \\ 0 & 0 & 1 & 1 \\ 0 & 0 & 0 & 1 \\ 1 & 0 & 0 & 0 \end{pmatrix}. \tag{10}$$

For large particle numbers and in the well-mixed limit the dynamics is described by Eqs. (1) in the main text and Eqs. (4), for direct competition and indirect competition, respectively.

We performed extensive stochastic simulations for the heterogeneous, asymmetric four species model. While the specific choice of $\mathbf{A}$ fixes the time scale, the reproduction rate for indirect competition was set to 1. Figure 7 shows the marginal distribution of each component of the genotypes for the model comprising direct competition at large times. Remarkably, as in the cyclic Lotka-Volterra model we find a striking influence of mobility on the evolution of phenotypic heterogeneity. This is most clear in the $P1$ and $S$ component. Interestingly, the transitions occur at the same values of the diffusion constant as in the heterogeneous, cyclic Lotka-Volterra model. By contrast, for indirect competition we find that phenotypic heterogeneity does not evolve in the asymmetric four species model. For large times, the population is dominated by specialists, focussing on one of the four phenotypic states. This is again in agreement with our findings for cyclic competition of three species.

## UNIVERSALITY OF SIMULATION RESULTS

Having found qualitatively the same results for two different ecological networks it seems reasonable to ask whether our findings are a universal property of a whole class of dynamic systems. Indeed, as outlined further above, the arguments provided in the previous sections are independent of the specific choice of the interaction matrix $\mathbf{A}$. Specifically, we expect to find the survival of bet-hedging in the well-mixed limit for any model comprising neutrally stable orbits in the homogeneous dynamics, and thereby a maximum number of conserved quantities (closed orbits) in the heterogeneous system [8]. To understand the different phases in the cyclic LV model we made use of a geometric interpretation of the net transitions rates, which argued for the monotonicity of $\overline{W}_\alpha^+$: specialists interact on faster time scales than generalists. In fact, we expect that these general arguments hold true for any model, where comprising neutrally stable orbits and a vanishing net interaction rate of genotypes with themselves and the bet-hedging genotype. We, therefore, believe that our findings on the evolution of phenotypic heterogeneity are not restricted to the models we studied here. Rather, we think that our results only depend on the basic properties of the underlying nonlinear dynamics.

The mechanisms responsible for the emergence of transitions in degree of mobility suggest that these transitions can not arise in less than two spatial dimensions. Indeed, simulations for one spatial dimension confirm that bet-hedgers dominate for all values of $D$ and direct competition [Fig. 8]. For indirect competition we again find the survival of specialists, regardless of the value of the diffusion constant.

## SAMPLE SIZES FOR CALCULATING AVERAGES AND HISTOGRAMS

For our results we made use of extensive stochastic simulations. Table I shows the number of simulation runs that have been performed in order to obtain averages and histograms.

TABLE I. Number of simulation runs used to calculate averages and histograms

| Figure | Number of runs |
|---|---|
| Main text | |
| Fig. 2 (a) | $10^5$ runs per value of $D$ |
| Fig. 2 (c) | $5 \cdot 10^4$ runs per value of $D$ |
| Fig. 3 (a), $D = 0$ | $6.8 \cdot 10^5$ |
| Fig. 3 (a), $D = 2 \cdot 10^{-5}$ | $9.6 \cdot 10^5$ |
| Fig. 3 (a), $D = 5 \cdot 10^{-5}$ | $7.4 \cdot 10^5$ |
| Fig. 3 (a), $D = 3 \cdot 10^{-3}$ | $6.6 \cdot 10^5$ |
| Fig. 3 (b) | $10^6$ |
| Fig. 3 (c) | $7.2 \cdot 10^5$ |
| Supplementary Material | |
| Fig. 1 (a) | $10^5$ runs per value of $N$ |
| Fig. 1 (b) | $5 \cdot 10^4$ per value of $N$ |
| Fig. 2 (c) | $6.8 \cdot 10^5$ |
| Fig. 5 | $10^7$ |
| Fig. 7 | $2.4 \cdot 10^5$ |
| Fig. 8 | $100$ |

[1] Note that the sum over all $\beta$ is actually zero and hence $\overline{W_\alpha^-} := \sum_{\beta \in \Delta^-} W_{\alpha\beta} = -\overline{W_\alpha^+}$.

[2] G. Szabó and T. Czárán, Phys. Rev. E **64**, 042902 (2001).

[3] G. Szabó, J. Phys. A: Math. Gen. **38**, 6689 (2005).

[4] P. Szabó, T. Czárán, and G. Szabó, J. Theor. Biol. **248**, 736 (2007).

[5] G. Szabó, A. Szolnoki, and I. Borsos, Phys. Rev. E **77**, 041919 (2008).

[6] T. Reichenbach, M. Mobilia, and E. Frey, Nature **448**, 1046 (2007).

[7] S. Rulands, A. Zielinski, and E. Frey, Phys. Rev. E **87**, 052710 (2013).

[8] J. Knebel, T. Krüger, M. F. Weber, and E. Frey, Phys. Rev. Lett. **110**, 168106 (2013).


\clearpage

\end{document}